\documentclass[prd,preprint,superscriptaddress,amsmath,amssymb,nofootinbib]{revtex4}
\usepackage{graphicx}
\usepackage{dcolumn}
\usepackage{bm}
\usepackage{amssymb}
\usepackage{amsmath}
\usepackage{epsfig}    
\usepackage{color}
\usepackage{slashed}
\usepackage{hhline}
\usepackage{youngtab}
\usepackage{multirow}

\makeatletter
\newcommand*\rel@kern[1]{\kern#1\dimexpr\macc@kerna}
\newcommand*\widebar[1]{%
  \begingroup
  \def\mathaccent##1##2{%
    \rel@kern{0.8}%
    \overline{\rel@kern{-0.8}\macc@nucleus\rel@kern{0.2}}%
    \rel@kern{-0.2}%
  }%
  \macc@depth\@ne
  \let\math@bgroup\@empty \let\math@egroup\macc@set@skewchar
  \mathsurround\z@ \frozen@everymath{\mathgroup\macc@group\relax}%
  \macc@set@skewchar\relax
  \let\mathaccentV\macc@nested@a
  \macc@nested@a\relax111{#1}%
  \endgroup
}
\makeatother

\def\be{\begin{equation}}
\def\ee{\end{equation}}
\newcommand{\bea}{\begin{eqnarray}}
\newcommand{\eea}{\end{eqnarray}}
\newcommand{\nn}{\nonumber}

\begin{document}

 \begin{flushright} EPHOU-23-017\\
KYUSHU-HET-270  \end{flushright}

\title{Modular flavor models with positive modular weights: \\ a new lepton model building}

\author{Tatsuo Kobayashi}
\email{kobayashi@particle.sci.hokudai.ac.jp}
\affiliation{Department of Physics, Hokkaido University, Sapporo 060-0810, Japan}

\author{Takaaki Nomura}
\email{nomura@scu.edu.cn}
\affiliation{College of Physics, Sichuan University, Chengdu 610065, China}

\author{Hiroshi Okada}
\email{okada.hiroshi@phys.kyushu-u.ac.jp}
\affiliation{Department of Physics, Kyushu University, 744 Motooka, Nishi-ku, Fukuoka, 819-0395, Japan}

\author{Hajime Otsuka}
\email{otsuka.hajime@phys.kyushu-u.ac.jp}
\affiliation{Department of Physics, Kyushu University, 744 Motooka, Nishi-ku, Fukuoka, 819-0395, Japan}

\date{\today}

\begin{abstract}
We propose an interesting assignment of positive modular wights for fields in a modular non-Abelian discrete flavor symmetry.
By this assignment, we can construct inverse seesaw and linear seesaw models without any additional symmetries which possess good testability in current experiments.
At first, we discuss probabilities for positive modular wights from a theoretical point of view.
Then we show concrete examples of inverse seesaw and linear seesaw scenarios applying modular $A_4$ symmetry as examples and
demonstrate some predictions as well as being consistent with experimental results such as their masses and mixings. 

 \end{abstract}
\maketitle

\section{Introduction}
Neutrinos may be Majorana type particles if we minimally extend the Standard Model (SM) by introducing right-handed (heavier) neutrinos.
This is called canonical seesaw scenario~\cite{Yanagida:1979gs, Minkowski:1977sc, Mohapatra:1979ia}. This scenario does not require any additional symmetry. Also, it leads us to an origin of Baryon number asymmetry via leptogenesis~\cite{Fukugita:1986hr}.
However this scenario typically requires a high energy scale for right-handed neutrino masses such as Grand Unified Theory (GUT) one ($\mathcal{O}(10^{16})$ GeV); this scale is also motivated to work the leptogensis well.
Thus it would be difficult to verify the canonical seesaw scenario directly by current/future experiments.
It is then important and interesting to explore other scenarios that could be tested. 

Inverse seesaw (IS)~\cite{Mohapatra:1986bd, Wyler:1982dd} and linear seesaw (LS)~\cite{Wyler:1982dd, Akhmedov:1995ip, Akhmedov:1995vm} mechanisms have been proposed as other possibilities of neutrino mass generation that enable us to obtain higher verifiability since these scenarios can induce the active neutrino masses with lower mass scale compared to canonical seesaw one, like TeV scale. However these models require an additional symmetry such as $U(1)_{B-L}$ to discriminate two types of heavier Majorana fermions. Also, there would not be any simple IS/LS models with predictive flavor structure which are nowadays tremendously important to verify them at collider experiments. This difficulty is because flavor structures cannot be constrained through such a new symmetry like $U(1)_{B-L}$. Thus we typically need multiple symmetries including flavor specific ones in realizing predictive IS/LS models. 
Phenomenologically, IS or LS model would be more attractive but a realization of predictive models is rather complicated than the canonical seesaw model because of the above reasons.

In this paper, we show an intriguing fact that one can construct minimal IS and LS models by introducing a modular $A_4$ symmetry only.
Moreover, due to determining the flavor structure for the lepton matter superfields by the symmetry, these models automatically possess high testability.
%
However, in realizing minimal models, we need to impose an unusual condition that we have to allow assignments with both plus and minus signs of the modular weight for four-dimensional (4D) superfields.
\footnote{When we only apply negative modular weights in a standard manner, we also need to introduce one more symmetry at least as we explain in Sec.III.}
Notice here that only the minus sign assignments for 4D fields are theoretically confirmed so far.
\footnote{Modular non-Abelian discrete flavor symmetries have recently been proposed in refs~\cite{Feruglio:2017spp, Criado:2018thu}. After that, a lot of models and their applications to flavor physics were studied. Here, we refer to models with modular $A_4$ symmetries.~\cite{Kobayashi:2018scp, Okada:2018yrn, Nomura:2019jxj, Okada:2019uoy, deAnda:2018ecu, Novichkov:2018yse, Nomura:2019yft, Okada:2019mjf,Ding:2019zxk, Nomura:2019lnr,Kobayashi:2019xvz,Asaka:2019vev,Zhang:2019ngf, Gui-JunDing:2019wap,Kobayashi:2019gtp,Nomura:2019xsb, Wang:2019xbo,Okada:2020dmb,Okada:2020rjb, Behera:2020lpd, Behera:2020sfe, Nomura:2020opk, Nomura:2020cog, Asaka:2020tmo, Okada:2020ukr, Nagao:2020snm, Okada:2020brs,Kang:2022psa, Nomura:2023usj}. See for more references Ref.~\cite{Kobayashi:2023zzc}.}
We thus need to investigate the theoretical verification of positive modular weights.
There exist some articles on quark and lepton masses and mixing assigning plus modular weights for fields under the $A_4$ symmetry. See for examples Refs.~\cite{Mishra:2023ekx, Liu:2021gwa}.

Here, we will discuss some theoretical possibilities on how to realize the positive signs of modular weight for 4D superfields. Then, we demonstrate two examples of neutrino models for IS and LS scenarios and display some predictions via numerical analyses. Our realization of these models indicates that the use of positive modular weights provides an extended possibility in constructing phenomenological models with modular flavor symmetry.

This paper is organized as follows. In Section~II, we present theoretical possibilities for positive modular weights for 4D fields.
In Section III, we show two examples of IS and LS scenarios on how to realize these models and demonstrate predictions via numerical analyses. 
Finally we devote Section~IV to the summary and conclusion of our results.

\section{Realizations of positive modular weights for chiral superfields from theoretical backgrounds}

In this section, we discuss the possibility of 4D positive modular weights in the context of higher-dimensional theory 
including the string theory.\footnote{See, Ref. \cite{Kikuchi:2022txy}, for the discussion of conventional 4D negative modular weights in the context of modular flavor models from a higher-dimensional perspective.} 
The 4D effective action can be obtained by using the Kaluza-Klein reduction on compact extra-dimensional spaces. 
For instance, higher-dimensional scalar fields $\Phi(x,y)$ and spinor fields $\Psi(x,y)$, with $x$ and $y$ being 
4D and $d$-dimensional internal coordinates, can be decomposed as
\begin{align}
    \Phi(x,y) = \sum_{n} \phi^{(n)}(x) \varphi^{(n)}(y),
    \nonumber\\
    \Psi(x,y) = \sum_{n} \psi^{(n)}(x) \chi^{(n)}(y),    
\end{align}
where $n$ labels Kaluza-Klein momentum in the internal space. 
Since the mass scale of Kaluza-Klein modes is typically a compactification scale, only massless modes with $n=0$ are 
relevant in the low-energy physics. 
Note that internal background sources lead to degenerate massless modes $\phi^{(0)}_i$, $\psi^{(0)}_i$ whose index $i$ will be 
identified with the generation number of quarks, leptons and/or Higgs. 
It indicates that the flavor structure of quarks and leptons will be determined by the structure of matter wavefunctions in extra-dimensional spaces $\varphi^{(0)}_i$, $\chi^{(0)}_i$. 
Hereafter, we omit the Kaluza-Klein index $n$ and flavor index $i$ unless we specify them. 

To see the flavor structure in more detail, let us discuss three-point couplings of matter wavefunctions. 
When higher-dimensional matter fields propagate in extra-dimensional spaces, 4D Yukawa couplings are given by 
an overlap integral of internal matter wavefunction such as $\varphi$ and $\chi$. 
For instance, the 4D Yukawa coupling term in the lepton sector is written by
\begin{align}
    W = Y \bar N H_u L, 
    \label{eq:4DW}
\end{align}
in the superfield language, where $\bar N$, $L$ and $H_u$ denote right-handed neutral fermions, left-handed leptons and up-sector 
Higgs field, respectively. 
Here, the 4D Yukawa coupling $Y$ is given by an overlap integral of corresponding matter wavefunctions in the extra-dimensional space:
\begin{align}
    Y = g \int d^d y \chi_{\bar N}(y) \chi_L(y) \varphi^\ast_{H_u}(y),
\end{align}
with $g$ being a higher-dimensional coupling. 
Remarkably, on certain toroidal backgrounds, such wavefunctions transform under finite modular groups $\Gamma_N = PSL(2,\mathbb{Z})/\Gamma(N)$ where $\Gamma(N)$ is the principal congruence subgroup:
\begin{align}
\begin{aligned}
\Gamma(N)= \left \{ 
\begin{pmatrix}
a & b  \\
c & d  
\end{pmatrix} \in SL(2,\mathbb{Z})~ ,
~~
\begin{pmatrix}
a & b  \\
c & d  
\end{pmatrix} =
\begin{pmatrix}
1 & 0  \\
0 & 1  
\end{pmatrix} ~~({\rm mod}\,N) \right \}
\end{aligned} .
\end{align}
Specifically, the modular transformation of internal wavefunction $\chi_i$ in its flavor space is given by
\begin{align}
    \chi_i(y) \rightarrow \rho(\gamma)_{ij} (c \tau +d)^k \chi_j(y),
\end{align}
where $\rho(\gamma)$ represents a reducible representation of $\Gamma_N$ and $k$ is the so-called modular weight.
Here, $\tau$ denotes the modulus of the torus. 
Similarly, the wavefunction $\varphi_i(y)$ transforms non-trivially.
It is notable that the higher-dimensional fields $\Phi(x,y)$ and $\Psi(x,y)$ themselves are invariant under the modular symmetry, indicating that the 4D matter fields $\phi_i(x)$ and $\psi_i(x)$ (or chiral superfields in the superfield language) enjoy opposite 
modular transformations in comparison with the corresponding wavefunctions $\varphi_i(y)$ and $\chi_i(y)$ in the extra-dimensional space, respectively. 
Thus, the 4D matters, $\phi_i(x)$ and $\psi_i(x)$, and internal fields,  $\varphi_i(y)$ and $\chi_i(y)$, have the modular weight 
$-k_i$ and $k_i$, respectively. 
The 4D superpotential (\ref{eq:4DW}) is indeed modular invariant because the 4D Yukawa coupling has the modular weight 
$k_Y = k_{\bar N} + k_{L} - k_{H_d}$.

The matter wavefunctions as well as the modular weights are explicitly calculated in toroidal backgrounds as well as its orbifolds. 
For instance, on $T^2$ background with magnetic fluxes, the modulus $\tau$ identified with the complex 
structure modulus determines the flavor structure of quarks and leptons. 
The massless wavefunction in the extra-dimensional space has the modular weight 1/2, and the corresponding Yukawa couplings are described by the modular form with weight 1/2 \cite{Kobayashi:2018rad,Kobayashi:2018bff,Ohki:2020bpo,Kikuchi:2020frp,Kikuchi:2020nxn,Kikuchi:2021ogn,Almumin:2021fbk}. 
Thus, the 4D matter fields have the modular weight $-1/2$.
(See for the modular symmetry and modular weights in heterotic orbifold models Refs. \cite{Lauer:1989ax,Lauer:1990tm,Ferrara:1989qb,Dixon:1989fj,Baur:2019kwi,Nilles:2020nnc,Nilles:2020gvu}.)
Such a negative modular weight of 4D matter fields is natural in the 4D effective action in the following sense. 
\begin{itemize}
    \item Property of modular forms from the viewpoint of higher-dimensional theories

    Firstly, suppose that the massless wavefunction in the extra-dimensional space $\varphi(y)$ is a holomorphic function of the modulus at the origin $y=0$.
    Then, when the wavefunction has the positive modular weight $k>0$, 
    it is a modular form. 
    Even if we incorporate the radiative corrections, the sign of modular weight will be unchanged as discussed in Ref. \cite{Kikuchi:2023clx}. 
    Indeed, in the opposite case $k<0$, $\varphi(0)$ will not be the modular form but rather a singular function. 

    \item Control of matter K\"ahler potential from the viewpoint of 4D effective theories

    Second, let us see the modular invariant K\"ahler potential of 4D matter fields:
\begin{align}
K =  \frac{1}{(2{\rm Im}(\tau ))^k}|\phi(x)|^2,
\label{eq:4DK}
\end{align}
corresponding to the normalization of internal wavefunction:
\begin{align}
\int d^dy\,\sqrt{g}\, |\varphi(y)|^2 = \frac{1}{(2{\rm Im}(\tau))^k}.
\end{align}
The positivity of modular weight $k$ controls higher-order corrections in the K\"ahler potential, 
for instance $|\phi(x)|^4/(2{\rm Im}(\tau ))^{2k}$, corresponding to the volume expansion of 
the torus. 
Hence, the negative modular weight will be out of control. 
    
\end{itemize}
Note that the above viewpoints are different from each other. The former point is to construct the 4D effective theories 
from fully known higher-dimensional theories by Kaluza-Klein reduction.
On the other hand, the latter point is the controllability of 
the 4D effective theories, which may be derived, e.g. by string 
perturbation theory of 4D modes. 

In the following, let us present two possibilities of realizing 4D positive modular weights from a top-down approach. 
\begin{enumerate}
    \item Localized modes

For the query of modular form, localized modes in extra dimensions may have a chance to have the positive modular weight for 4D chiral matters. 
Suppose that they are localized at $y\neq 0$.
Then, their wavefunctions vanish at $y=0$.
These modes may be free from the discussion on modular forms.
For example, the heterotic string theory on toroidal orbifolds has untwisted and twisted modes. 
The twisted modes are localized modes, while the untwisted modes are bulk modes.
The ground states of twisted modes on $T^2/Z_N$ have the modular weights $-1/N$ for 4D matter fields. 
On top of that, the oscillators shift the modular weight by $\pm 1$, where the sign depends on the property of oscillators. 
Thus, 4D matter fields have a chance of the positive modular weights. (See, Refs. \cite{Dixon:1989fj,Ibanez:1992hc,Kawabe:1994mj}, for an explicit construction.) 
The matter K\"ahler potential would be controlled in the stringy calculations or the following multi moduli scenario.

    \item Bulk modes in a multi moduli scenario

For the query of controlling the K\"ahler potential, multi moduli scenario will be crucial. 
For instance, in the two moduli case, the matter K\"ahler potential will be expanded in the large volume regime:
\begin{align}
K =  \frac{1}{(2{\rm Im}(T))^n(2{\rm Im}(t))^k}|\phi(x)|^2,
\label{eq:4DK_multi}
\end{align}
where $n$ and $k$ are modular weights associated with the modulus $T$ and $t$, respectively. 
Let us suppose that the modular flavor symmetry is originated from the modulus $t$, but the overall volume 
is determined by another modulus $T$. 
In this case, the negative modular weight $k<0$ will be controllable. 
For illustrative purposes, we deal with the $E_8\times E_8$ heterotic string with standard embedding, where the matter K\"ahler potential will be extracted from the moduli K\"ahler potential. 
Specifically, on $T^6/\mathbb{Z}_3$ orbifold, the K\"ahler potential is known up to the second order in the chiral matter field $A_\alpha$ which is a ${\bf 27}$ fundamental representation of $E_6$ \cite{Ferrara:1986qn,Cvetic:1988yw}:\footnote{A hierarchical structure of holomorphic Yukawa couplings was discussed on this background \cite{Ishiguro:2021drk}.}
\begin{align}
    K &= -\ln \det (2{\rm Im}(T)-A^\alpha A^{\bar{\alpha}})^{a\bar{b}} 
    = -\ln \det (2{\rm Im}(T))^{a\bar{b}} + (2{\rm Im}(T))^{-1}_{a\bar{b}}A^a_\alpha A^{\bar{b}}_{\bar{\alpha}},
\end{align}
with 
\begin{align}
T=
\begin{pmatrix}
T^1 & T^4 & T^5\\
T^7 & T^2 & T^6\\
T^8 & T^9 & T^3\\
\end{pmatrix}
.
\end{align}
Here, $a$ and $\alpha$ denote the index of moduli and the index of $SU(3)\subset E_8$, respectively. 
Note that we focus on untwisted K\"ahler moduli and corresponding untwisted matters. 
When we restrict ourselves on the locus $T:= T^1=T^2=T^3$ and $t:=T^4=T^5=T^6=T^7=T^8=T^9$, an explicit form of matter K\"ahler metric $K_{a\bar{b}}^{(27)}=(2{\rm Im}(T))^{-1}_{a\bar{b}}$ is
\begin{align}
&2({\rm Im}(T) - {\rm Im}(t))^2({\rm Im}(T) + 2{\rm Im}(t))K_{a\bar{b}}^{(27)}
\nonumber\\
&=
\begin{pmatrix}
{\rm Im}(T)^2 - {\rm Im}(t)^2 & 
({\rm Im}(t) - {\rm Im}(T)) {\rm Im}(t) & 
({\rm Im}(t) - {\rm Im}(T)) {\rm Im}(t) \\
({\rm Im}(t) - {\rm Im}(T)) {\rm Im}(t) & 
{\rm Im}(T)^2 - {\rm Im}(t)^2 & 
({\rm Im}(t) - {\rm Im}(T)) {\rm Im}(t) \\
({\rm Im}(t) - {\rm Im}(T)) {\rm Im}(t) &
({\rm Im}(t) - {\rm Im}(T)) {\rm Im}(t) &
{\rm Im}(T)^2 - {\rm Im}(t)^2 \\
\end{pmatrix}
,
\label{eq:KahlerZ3}
\end{align}
\normalsize
which can be diagonalized as
\begin{align}
\Lambda_{\hat{a}\bar{\hat{b}}} =&
\begin{pmatrix}
(2{\rm Im}(T) -2{\rm Im}(t))^{-1} & 
0 & 
0 \\
0 & 
(2{\rm Im}(T) -2{\rm Im}(t))^{-1} & 
0 \\
0 &
0 &
(2{\rm Im}(T) +4{\rm Im}(t))^{-1}\\
\end{pmatrix}
.
\label{eq:LambdaZ3iso}
\end{align}
Thus, when we focus on the $t$-dependent K\"ahler metric in the diagonal basis, 
one can arrive at the matter K\"ahler metric with 4D positive modular weight:
\begin{align}
    K &\propto  \frac{2{\rm Im}(t)}{2{\rm Im}(T)}|\phi(x)|^2,
\end{align}
where it is notable that the overall volume is now controlled by ${\rm Im}(T)$ rather than ${\rm Im}(t)$, and both of them are considered to be larger than the string length. 
It results in the control of matter K\"ahler potential against higher-order corrections. 
A moderate hierarchy between the overall volume and the volume of local cycle plays an important role in the realization of positive modular weights. 
Since the modular flavor symmetry is realized in the local cycle of underlying manifolds, the matter wavefunctions would be localized in the extra-dimensional space. Thus, the query of modular form would be resolved, but it is required to check the property of wavefunctions themselves, which left for future work. 
This is just a specific example of realizing the 4D positive modular weight for the untwisted sector on toroidal orbifolds, but we 
expect similar phenomena in a more generic multi moduli scenario. 
Indeed, on the so-called ``Swiss-cheese like Calabi-Yau manifolds" as employed in the Large Volume Scenario \cite{Balasubramanian:2005zx,Conlon:2005ki}, 
one can derive a similar structure of moduli K\"ahler potential due to the hierarchy between the overall volume and the local divisor volume. 
Following a general procedure for constructing modular flavor models on Calabi-Yau manifolds \cite{Ishiguro:2020nuf,Ishiguro:2021ccl}, we would realize the 4D positive modular weights whose comprehensive study will be left for future work.

\end{enumerate}

\section{Two examples of lepton models}

In this section, we show two examples of lepton models: ``Inverse Seesaw (IS)" and ``Linear Seesaw (LS)", introducing positive modular weights. Thanks to the extension of degrees of freedom for signs of modular weights, we find that these two scenarios can be constructed by this symmetry itself.~\footnote{ 
Notice here that we implicitly work on the supersymmetric theory to forbid infinite terms by making good use of holomorphicity from supersymmetry, and we start from its effective theory after breaking supersymmetry. See, Refs. \cite{Kobayashi:2021uam,Kikuchi:2022pkd}, for a realization of modular-invariant effective field theory after the soft supersymmetry breaking.} 
Without positive and negative modular weights, 
these two scenarios can be constructed only if additional symmetry such as a $U(1)$ symmetry is imposed. 
This is (roughly speaking) because we have to differentiate two types of heavier Majorana fermions denoted by $\bar N$ and $S$, and this discrimination cannot be achieved by applying negative modular weights only.
Below we will show how to realize our models.

Before discussing each of the neutrino mass mechanisms, we review general formulation on neutrino masses and their mixings.
The neutrino mass matrix can be written in terms of overall mass dimension parameter $\kappa$ and dimensionless neutrino mass matrix $\tilde m_\nu$:
\begin{align}
m_\nu \equiv \kappa \tilde m_\nu,
\end{align}
where $\kappa$ and $\tilde m_\nu$ are determined if a concrete mechanism is fixed.
$m_\nu$ is diagonalized by a unitary matrix $V_{\nu}$; $D_\nu=|\kappa| \tilde D_\nu= V_{\nu}^T m_\nu V_{\nu}=|\kappa| V_{\nu}^T \tilde m_\nu V_{\nu}$. 
Then $|\kappa|$ is fixed by
\begin{align}
(\mathrm{NH}):\  |\kappa|^2= \frac{|\Delta m_{\rm atm}^2|}{\tilde D_{\nu_3}^2-\tilde D_{\nu_1}^2},
\quad
(\mathrm{IH}):\  |\kappa|^2= \frac{|\Delta m_{\rm atm}^2|}{\tilde D_{\nu_2}^2-\tilde D_{\nu_3}^2},
 \end{align}
where $\Delta m_{\rm atm}^2$ is the atmospheric neutrino mass-squared splitting, and NH and IH indicate the normal hierarchy and the inverted hierarchy, respectively. 
Subsequently, the solar mass squared splitting can be obtained in terms of $|\kappa|$ such that:
\begin{align}
\Delta m_{\rm sol}^2=  |\kappa|^2 ({\tilde D_{\nu_2}^2-\tilde D_{\nu_1}^2}),
 \end{align}
 which is compared to the observed value in our numerical analysis.
 %
We define the observed mixing matrix by $U=V^\dag_{eL} V_\nu$~\cite{Maki:1962mu}~\footnote{Note that if the charged-lepton mass matrix $m_\ell$ is not a diagonal one, we need to diagonalize it as ${\rm  diag}( |m_e|^2, |m_\mu|^2, |m_\tau|^2)\equiv V_{e_L}^\dag m^\dag_\ell m_\ell V_{e_L}$, where $V_{eL}$ is a unitary matrix.}, that
is parametrized by three mixing angles $\theta_{ij} (i,j=1,2,3; i < j)$, one CP violating Dirac phase $\delta_{CP}$,
and two Majorana phases $\alpha_{21},\alpha_{31}$.
Under the parametrization, we write the matrix by
\begin{equation}
U = 
\begin{pmatrix} c_{12} c_{13} & s_{12} c_{13} & s_{13} e^{-i \delta_{CP}} \\ 
-s_{12} c_{23} - c_{12} s_{23} s_{13} e^{i \delta_{CP}} & c_{12} c_{23} - s_{12} s_{23} s_{13} e^{i \delta_{CP}} & s_{23} c_{13} \\
s_{12} s_{23} - c_{12} c_{23} s_{13} e^{i \delta_{CP}} & -c_{12} s_{23} - s_{12} c_{23} s_{13} e^{i \delta_{CP}} & c_{23} c_{13} 
\end{pmatrix}
\begin{pmatrix} 1 & 0 & 0 \\ 0 & e^{i \frac{\alpha_{21}}{2}} & 0 \\ 0 & 0 &  e^{i \frac{\alpha_{31}}{2}} \end{pmatrix},
\end{equation}
where $c_{ij}$ and $s_{ij}$ stand for $\cos \theta_{ij}$ and $\sin \theta_{ij}$ ($i,j=1-3$), respectively. 
Then, we can write each of the mixing in terms of the component of $U$ as follows:
\begin{align}
\sin^2\theta_{13}=|U_{e3}|^2,\quad 
\sin^2\theta_{23}=\frac{|U_{\mu3}|^2}{1-|U_{e3}|^2},\quad 
\sin^2\theta_{12}=\frac{|U_{e2}|^2}{1-|U_{e3}|^2}.
\end{align}
The Dirac phase $\delta_{CP}$ is also given by the Jarlskog invariant:
\begin{align}
\sin \delta_{CP} &= \frac{\text{Im} [U_{e1} U_{\mu 2} U_{e 2}^* U_{\mu 1}^*] }{s_{23} c_{23} s_{12} c_{12} s_{13} c^2_{13}} ,\quad
\cos \delta_{CP} = -\frac{|U_{\tau1}|^2 -s^2_{12}s^2_{23}-c^2_{12}c^2_{23}s^2_{13}}{2 c_{12} s_{12} c_{23} s_{23}s_{13}} ,
\end{align}
where $\delta_{CP}$ being subtracted from $\pi$ if $\cos \delta_{CP}$ is negative.
In addtion Majorana phase $\alpha_{21},\ \alpha_{31}$ are found from following relations:
\begin{align}
&
\sin \left( \frac{\alpha_{21}}{2} \right) = \frac{\text{Im}[U^*_{e1} U_{e2}] }{ c_{12} s_{12} c_{13}^2} ,\quad
  \cos \left( \frac{\alpha_{21}}{2} \right)= \frac{\text{Re}[U^*_{e1} U_{e2}] }{ c_{12} s_{12} c_{13}^2}, \
%
,\\
&
 \sin \left(\frac{\alpha_{31}}{2}  - \delta_{CP} \right)=\frac{\text{Im}[U^*_{e1} U_{e3}] }{c_{12} s_{13} c_{13}},
\quad 
 \cos \left(\frac{\alpha_{31}}{2}  - \delta_{CP} \right)=\frac{\text{Re}[U^*_{e1} U_{e3}] }{c_{12} s_{13} c_{13}},
\end{align}
where $\alpha_{21}/2,\ \alpha_{31}/2-\delta_{CP}$
are subtracted from $\pi$, when $ \cos \left( \frac{\alpha_{21}}{2} \right),\  \cos \left(\frac{\alpha_{31}}{2}  - \delta_{CP} \right)$ values are negative.
Finally, the effective mass for the neutrinoless double beta decay is given by
\begin{align}
\langle m_{ee}\rangle=|\kappa||\tilde D_{\nu_1} \cos^2\theta_{12} \cos^2\theta_{13}+\tilde D_{\nu_2} \sin^2\theta_{12} \cos^2\theta_{13}e^{i\alpha_{21}}+\tilde D_{\nu_3} \sin^2\theta_{13}e^{-2i\delta_{CP}}|,
\end{align}
where its predicted value could be tested by current/future experiments such as KamLAND-Zen~\cite{KamLAND-Zen:2016pfg}.

\subsection{Inverse seesaw}
\begin{center} 
\begin{table}[tb]
\begin{tabular}{|c||c|c|c|c||c|c||}\hline\hline  
&\multicolumn{4}{c||}{ Leptons} & \multicolumn{2}{c||}{Higgs} \\\hline
  & ~$(L_{e}$,$L_{\mu}$,$L_{\tau})$~&~ ($\bar e$,$\bar \mu$,$\bar \tau$)~ & ~$\bar N$~ & ~$S$~ & ~$H_u$~ & ~$H_d$~
  \\\hline 
 $SU(2)_L$ & $\bm{2}$  & $\bm{1}$  & $\bm{1}$ & $\bm{1}$   & $\bm{2}$  & $\bm{2}$   \\\hline 
$U(1)_Y$ & $-\frac12$ & $+1$ & $0$  & $0$& $\frac12$ & $-\frac12$     \\\hline
 $A_4$ & $\{1\}$ & $\{\bar1\}$ & $\{\bar1\}$ & $3$ & $1$ & $1$   \\\hline
 $-k$ & $+1$ & $-1$ & $-1$ & $-1$ & $0$ & $0$     \\\hline
\end{tabular}
\caption{(IS):
Field contents for leptons and Higgs
with their charge assignments under $SU(2)_L\otimes U(1)_Y \otimes A_4$ where $-k$ is the number of modular weight and $\{1\}\equiv \{1,1',1''\}$ and $\{\bar1\}\equiv \{1,1'',1'\}$.}
\label{tab:is}
\end{table}
\end{center}

The IS scenario requires two types of neutral fermions $\bar N$ and $S$ where we assume each of them has three families.
Our model is discriminated by different charges of flavor symmetry of $A_4$; three types of singlets $\{\bar1\}\equiv\{1,1'',1'\}$ for $\bar N$ and triplet $3$ for $S$ where $-1$ modular weights are imposed for all these neutral fermions.
The SM leptons are assigned under $A_4$ as follows:
$\{\bar1\}$ for three right-handed charged-leptons $\{\bar e,\bar \mu,\bar\tau\}$ and $\{1\}(\equiv\{1,1',1''\})$ for three left-handed leptons $\{L_e,L_\mu,L_\tau\}$ where $+1$ and $-1$ modular weights are imposed respectively. Due to these assignments, we find the charged-lepton mass matrix is diagonal at the Lagrangian. Therefore, the mass eigenvalues of charged-leptons are given by $m_\ell\equiv y_\ell v_d/\sqrt2$ ($\ell\equiv(e,\mu,\tau)$) where $\langle H_d\rangle\equiv [0,v_d/\sqrt2]^T$.
It implies that PMNS matrix denoted by $U$ directly comes from the neutrino sector.
Thus, we concentrate on the neutrino sector below.
Moreover, we introduce two Higgs fields to cancel chiral anomalies 
as the usual supersymmetric SM.
We summarize all the charge assignments and their field contents in Tab.~\ref{tab:is}.
Under these symmetries, renormalizable superpotential is found as
\begin{align}
W_{(IS)}&= \sum_{i=1}^3 y_{D_i} \bar N_i H_u L_i \\
&+
m_{NS} \left[\bar N_1 (y_1 S_1 + y_3 S_2+ y_2 S_3)
+
\tilde\alpha_2 \bar N_2 (y_2 S_1 + y_1 S_2+ y_3 S_3)
+
\tilde\alpha_3 \bar N_3 (y_3 S_1 + y_2 S_2+ y_1 S_3)\right] \nn\\
&+m_S \left[y_1(2S_1S_1-S_2S_3-S_3S_2) +y_2(2S_2S_2-S_1S_3-S_3S_1)
+y_3(2S_3S_3-S_1S_2-S_2S_1) \right],\nn
\end{align}
where $y_{1,2,3}$ is Yukawa coupling given by a modular form under $A_4$ triplet with modular weight $2$, and all the other terms are forbidden by our charge assignments.
In the lepton sector at least, R-parity is not needed because of the modular $A_4$ symmetry.

Notice here that $\bar N H_u L$ term is absent if we choose $-1$ modular weight for $L_{i}$ since $A_4$ triplet weight 2 modular form is required while superfields are singlet. It is indeed possible to have the term assigning a triplet for $\bar N$ but we can also have $L S H_u$ and $\bar N \bar N$ terms breaking the IS structure. Also if we choose $-(2n+1)$\,($n \neq 0$) weight for $L_i$, we generally have $L S H_u$ term. In general, we can not obtain the IS structure in a minimal way with negative modular weights only; there will be more possible invariant terms if we increase the absolute value of modular weights or assign triplets for other superfields. Therefore positive modular weights make it possible to realize the IS structure without imposing any other symmetry.

 After spontaneous symmetry breaking, we find  nine by nine neutral fermion's mass matrix in a basis of $[\nu,\bar N^C, S]^T$ as follows
 \begin{align}
 M_N\equiv  
\begin{pmatrix} 0 & m_D^T & 0 \\
m_D & 0 & M_{NS}\\
0 & M_{NS}^T & M_S\end{pmatrix},
 \end{align}
 where $m_D, M_S, M_{{NS}}$ are respectively given by
\begin{align}
m_D&\equiv \frac{y_{D_1} v_u}{\sqrt2} \tilde m_D ,\quad
\tilde m_D\equiv  
\begin{pmatrix} 1 & 0 & 0 \\
0 &\tilde\mu_2 & 0\\
0 & 0 &\tilde\mu_3
\end{pmatrix}, \\
\mu_{S}&\equiv m_S \tilde \mu_{S} ,\quad
\tilde \mu_{S} \equiv  
\begin{pmatrix} 2 y_1 & -y_3 & -y_2 \\
-y_3 & 2y_2 & -y_1\\
-y_2 & -y_1 &2 y_3
\end{pmatrix}, \\
M_{NS}&\equiv m_{NS} \tilde M_{NS} ,\quad
\tilde M_{NS} \equiv  
\begin{pmatrix} 1 & 0 & 0 \\
0 & \tilde\alpha_2 & 0\\
0 & 0 & \tilde\alpha_3
\end{pmatrix}
\begin{pmatrix} y_1 & y_3 & y_2 \\
y_2 & y_1 & y_3\\
y_3 & y_2 &y_1
\end{pmatrix},
\end{align}
 and $\langle H_u\rangle\equiv [v_u/\sqrt2,0]^T$.
 Here, all the dimensionless parameters (except $y_{1,2,3}$); $\tilde \mu_{2,3}, \tilde \alpha_{2,3}$, can be real after rephasing of fields.
Once we assume the following mass hierarchy $\mu_S \ll m_D<M_{NS}$,
we find the active neutrino mass matrix in light of the IS mechanism as follows:
\begin{align}
m_\nu \approx (m_D M_{ND}^{-1}) \mu_S (m_D M_{ND}^{-1})^T.
\end{align}
Furthermore, $m_\nu$ can be rewritten in terms of dimensionless matrices:
\begin{align}
m_\nu \approx \frac{y_{D_1}^2 v_u^2 m_S}{2 m^2_{NS}}(\tilde m_D \tilde M_{ND}^{-1}) \tilde \mu_S (\tilde m_D \tilde M_{ND}^{-1})^T\equiv \kappa \tilde m_\nu,
\end{align}
where $\kappa\equiv \frac{y_{D_1}^2 v_u^2 m_S}{2 m^2_{NS}}$ and $\tilde m_\nu\equiv (\tilde m_D \tilde M_{ND}^{-1}) \tilde \mu_S (\tilde m_D \tilde M_{ND}^{-1})^T$.
Now we have four dimensionless real parameters $\tilde \mu_{2,3}, \tilde \alpha_{2,3}$ and one complex one $\tau$. Therefore, we totally have six free parameters.

\subsection{Linear seesaw}
\begin{center} 
\begin{table}[tb]
\begin{tabular}{|c||c|c|c|c||c|c||}\hline\hline  
&\multicolumn{4}{c||}{ Leptons} & \multicolumn{2}{c||}{Higgs} \\\hline
  & ~$(L_{e}$,$L_{\mu}$,$L_{\tau})$~&~ ($\bar e$,$\bar \mu$,$\bar \tau$)~ & ~$\bar N$~ & ~$S$~ & ~$H_u$~ & ~$H_d$~
  \\\hline 
 $SU(2)_L$ & $\bm{2}$  & $\bm{1}$  & $\bm{1}$ & $\bm{1}$   & $\bm{2}$  & $\bm{2}$   \\\hline 
$U(1)_Y$ & $-\frac12$ & $+1$ & $0$  & $0$& $\frac12$ & $-\frac12$     \\\hline
 $A_4$ & $3$ & $\{\bar1\}$ & $\{\bar1\}$ & $\{1\}$ & $1$ & $1$   \\\hline
 $-k$ & $-3$ & $-1$ & $+1$ & $-1$ & $0$ & $0$     \\\hline
\end{tabular}
\caption{(LS):
Field contents for leptons and Higgs
with their charge assignments under $SU(2)_L\otimes U(1)_Y\otimes U(1)_{H} \otimes A_4$ where $-k$ is the number of modular weight and $\{1\}\equiv \{1,1',1''\}$ and $\{\bar1\}\equiv \{1,1'',1'\}$.}
\label{tab:ls}
\end{table}
\end{center}

The LS scenario also requires two types of neutral fermions $\bar N$ and $S$ where we assume each of them has three families.
Our model is discriminated by different charges of modular flavor symmetry of $A_4$; three types of singlets $\{\bar1\}$ with $+1$ modular weight for $\bar N$, and singlets $\{1\}$ for three generations of $S$ where $-1$ modular weights are imposed for them.
The SM leptons are assigned under $A_4$ as follows:
$\{\bar1\}$ for three right-handed charged-leptons $\{\bar e,\bar \mu,\bar\tau\}$ and triplet for three left-handed leptons $\{L_e,L_\mu,L_\tau\}$ where $-1$ and $-3$ modular weights are imposed respectively. 
The Higgs sector is the same as the one in IS case.
We summarize all the charge assignments and their field contents in Tab.~\ref{tab:ls}.
Under these symmetries, renormalizable superpotential is found as
\begin{align}
W_{(LS)}&= 
a_e \bar e H_d (y'_1 L_1+y'_3 L_2+y'_2 L_3) 
+
b_e \bar \mu H_d (y'_2 L_1+y'_1 L_2+y'_3 L_3) \nn\\
&+
c_e \bar \tau H_d (y'_3 L_1+y'_2 L_2+y'_1 L_3)
\\
&+\alpha_{D_1}\left[\bar N_1 H_u (y_1 L_1+y_3 L_2+y_2 L_3) 
+
\tilde\alpha_{D_2} \bar N_2 H_u (y_2 L_1+y_1 L_2+y_3 L_3) \right.\nn\\
&+
\left.\tilde\alpha_{D_3} \bar N_3 H_u (y_3 L_1+y_2 L_2+y_1 L_3) \right]
\\
&+\alpha_{D'_1}\left[S_1 H_u (y'_1 L_1+y'_3 L_2+y'_2 L_3) 
+
\tilde\alpha_{D'_2} S_2 H_u (y'_3 L_1+y'_2 L_2+y'_1 L_3) \right.\nn\\
&+
\left.\tilde\alpha_{D'_3} S_3 H_u (y'_2 L_1+y'_1 L_2+y'_3 L_3) \right]
\\
&+
\sum_{i=1}^3 m_{NS_i}\bar N_i S_i,
\end{align}
where $y'_{1,2,3}$ is Yukawa coupling given by modular form under $A_4$ triplet with modular weight $4$, and all the other terms are forbidden by our charge assignments.

Notice here that $\bar N S$ term is absent if we choose $-1$ modular weight for $\bar N$. It is possible to get the term by considering $-(2n+1)$\,($n \neq 0$) weight for $\bar N$ but this weight also allows $\bar N \bar N^T$ term. Therefore positive modular weight is crucial to realize LS structure in a minimal way as in the IS case.

Charged-lepton mass matrix is given after spontaneous symmetry breaking of $H_d$:
\begin{align}
m_\ell = \frac{v_d}{\sqrt2}
\begin{pmatrix} 
a_e & 0 & 0 \\
0 & b_e & 0\\
0 & 0 & c_e
\end{pmatrix}
\begin{pmatrix} 
y'_1 & y'_3 & y'_2 \\
y'_2 & y'_1 & y'_3\\
y'_3 & y'_2 & y'_1\end{pmatrix},
 \end{align}
 where $a_e,b_e,c_e$ are real without loss of generality.
 Then the charged-lepton mass eigenvalues are found as ${\rm  diag}( |m_e|^2, |m_\mu|^2, |m_\tau|^2)\equiv V_{e_L}^\dag m^\dag_\ell m_\ell V_{e_L}$. We fix these three input parameters by inserting the observed charged-lepton masses and $V_{e_L}$ as follows:
\begin{align}
&{\rm Tr}[m_\ell {m_\ell}^\dag] = |m_e|^2 + |m_\mu|^2 + |m_\tau|^2,\\
&{\rm Det}[m_\ell {m_\ell}^\dag] = |m_e|^2  |m_\mu|^2  |m_\tau|^2,\\
&({\rm Tr}[m_\ell {m_\ell}^\dag])^2 -{\rm Tr}[(m_\ell {m_\ell}^\dag)^2] =2( |m_e|^2  |m_\mu|^2 + |m_\mu|^2  |m_\tau|^2+ |m_e|^2  |m_\tau|^2 ).
\end{align}
Here, we apply the experimental values summarized in PDG for the charged-lepton masses~\cite{ParticleDataGroup:2018ovx}.
\\
 After spontaneous symmetry breaking, we find  nine by nine neutral fermion's mass matrix in a basis of $[\nu,\bar N^C, S]^T$ as follows
 \begin{align}
 M_N\equiv  
\begin{pmatrix} 0 & m_D^T & m'^T_D \\
m_D & 0 & M_{NS}\\
m'_D & M_{NS}^T & 0\end{pmatrix}
+
\begin{pmatrix} 0 & m_D^T & m'^T_D \\
m_D & 0 & M_{NS}\\
m'_D & M_{NS}^T & 0\end{pmatrix}^T,
 \end{align}
 where $m_D, M_S, M_{{NS}}$ are respectively given by
\begin{align}
m_D&\equiv \frac{\alpha_{D_1} v_u}{\sqrt2} \tilde m_D ,\quad
\tilde m_D\equiv  
\begin{pmatrix} 1 & 0 & 0 \\
0 &\tilde\alpha_{D_2} & 0\\
0 & 0 &\tilde\alpha_{D_3}
\end{pmatrix}
\begin{pmatrix} y_1 & y_3 & y_2 \\
y_2 & y_1 & y_3\\
y_3 & y_2 &y_1
\end{pmatrix}, \\
m'_D&\equiv \frac{\alpha_{D'_1} v_u}{\sqrt2} \tilde m'_D ,\quad
\tilde m'_D \equiv  
\begin{pmatrix} 1 & 0 & 0 \\
0 &\tilde\alpha_{D'_2} & 0\\
0 & 0 &\tilde\alpha_{D'_3}
\end{pmatrix}
\begin{pmatrix} y'_1 & y'_3 & y'_2 \\
y'_3 & y'_2 &y'_1\\
y'_2 & y'_1 & y'_3
\end{pmatrix}, \\
M_{NS}&\equiv m_{NS_1} \tilde M_{NS} ,\quad
\tilde M_{NS} \equiv  
\begin{pmatrix} 1 & 0 & 0 \\
0 & \tilde\mu_{NS_2} & 0\\
0 & 0 & \tilde\mu_{NS_3}
\end{pmatrix},
\end{align}
 $\tilde\mu_{NS_a}\equiv m_{NS_a}/m_{NS_1}(a=2,3)$, and $\langle H_u\rangle\equiv [v_u/\sqrt2,0]^T$.
 Here,  four dimensionless parameters $\tilde \alpha_{D_{2,3}},\tilde \alpha_{D'_{2,3}}$ can be real after rephasing of fields while $\tilde \mu_{NS_{2,3}}$ are complex.
Once we assume the following mass hierarchy $m_D,m'_D \ll M_{NS}$,
we find the active neutrino mass matrix in light of the LS mechanism as follows:
\begin{align}
m_\nu \approx m'^T_D M_{ND}^{-1} m_D^T + m_D (M^T_{ND})^{-1}m'_D.
\end{align}
Furthermore, $m_\nu$ can be rewritten in terms of dimensionless matrices:
\begin{align}
m_\nu \approx \frac{\alpha_{D_1}\alpha_{D'_1} v_u^2}{2 m_{NS_1}}
[\tilde m'^T_D \tilde M_{ND}^{-1} \tilde m_D^T + \tilde m_D (\tilde M^T_{ND})^{-1}\tilde m'_D]\equiv \kappa \tilde m_\nu,
\end{align}
where $\kappa\equiv \frac{\alpha_{D_1}\alpha_{D'_1} v_u^2}{2 m_{NS_1}}$ and $\tilde m_\nu\equiv \tilde m'^T_D \tilde M_{ND}^{-1} \tilde m_D^T + \tilde m_D (\tilde M^T_{ND})^{-1}\tilde m'_D$.
Now we have four dimensionless real parameters $\tilde \alpha'_{2,3}, \tilde \alpha_{2,3}$, and three complex ones $\tau, \tilde \mu_{NS_{2,3}}$. Therefore, we totally have ten free parameters.

\subsection{Numerical analysis}
We have numerical analysis for IS and LS models below and show what kind of correlations are found.
{
Before showing the numerical analysis, we need to discuss a constraint from non-unitarity that should be taken in account for IS and LS. The non-unitarity matrix denoted by $U'$ that represents the deviation from the unitarity, and  it is typically parametrized by the following form 
\begin{align}
U'\equiv \left(1-\frac12 FF^\dag\right) U,
\end{align}
where $F\equiv  (M^*_{NS})^{-1} m^T_D$ is a hermitian matrix.
$M_{NS}$ and $m_D$ are the same notations as the ones of each model for IS and LS.
The global constraints are found via several experimental results such as the SM $W$ boson mass, the effective Weinberg angle, several ratios of $Z$ boson fermionic decays, invisible decay of $Z$, electroweak universality, measured Cabbibo-Kobayashi-Maskawa, and lepton flavor violations~\cite{Fernandez-Martinez:2016lgt}.
These results give the following constraints for $|FF^\dag|$~\cite{Agostinho:2017wfs}
\begin{align}
|FF^\dag|\le  
\left[\begin{array}{ccc} 
2.5\times 10^{-3} & 2.4\times 10^{-5}  & 2.7\times 10^{-3}  \\
2.4\times 10^{-5}  & 4.0\times 10^{-4}  & 1.2\times 10^{-3}  \\
2.7\times 10^{-3}  & 1.2\times 10^{-3}  & 5.6\times 10^{-3} \\
 \end{array}\right].
\end{align} 
In our numerical analyses below, 
we implicitly impose the above constraints in addition to the neutrino oscillation data.

\subsubsection{IS}
We randomly select our free parameters within the following range:
\begin{align}
\{\tilde\alpha_{2,3},\tilde\mu_{2,3}\}\in [10^{-3} -10^3],
\end{align}
and we work on the fundamental region on $\tau$.

\begin{figure}[tb]
\begin{center}
\includegraphics[width=77.0mm]{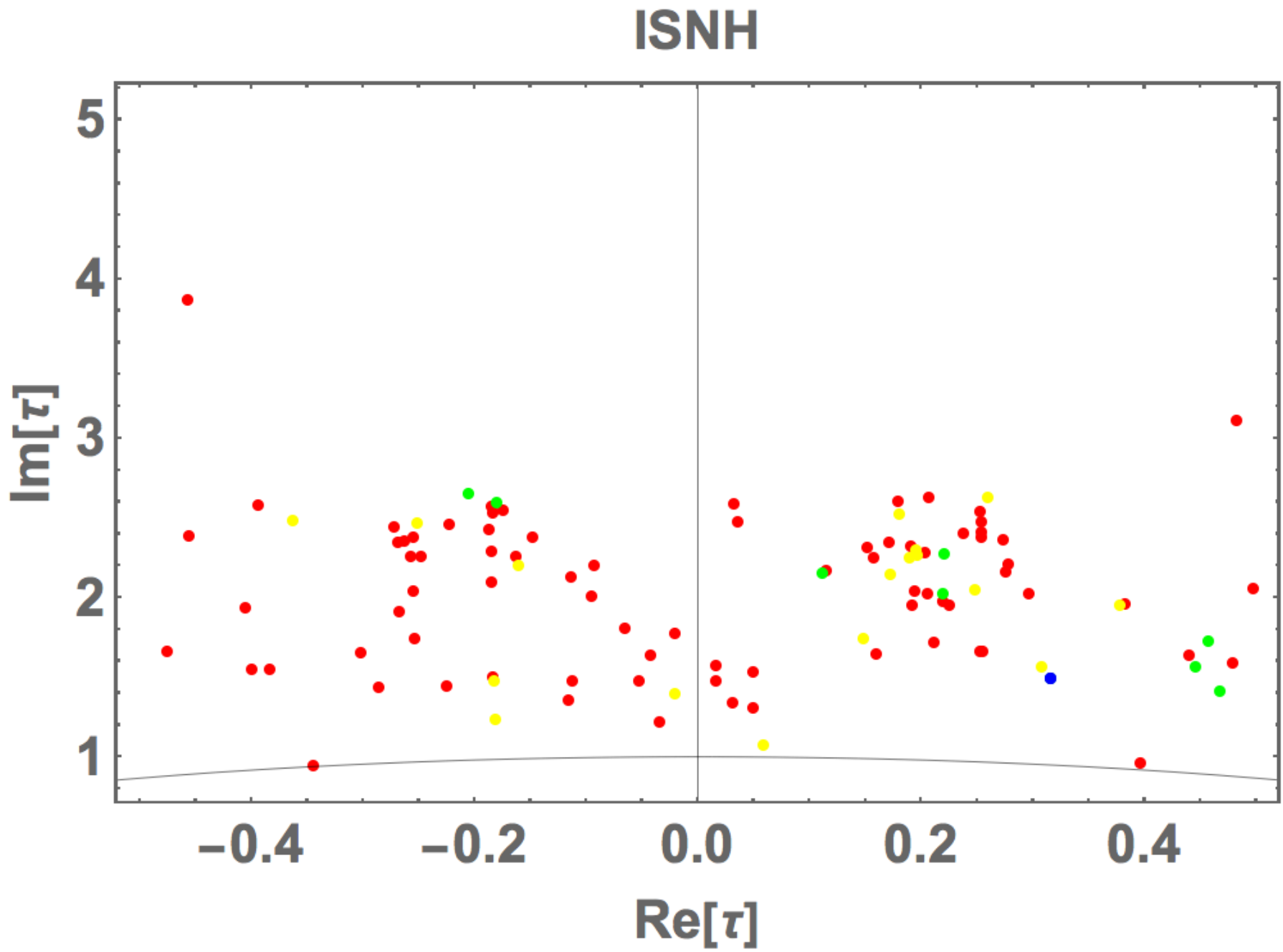}
\includegraphics[width=77.0mm]{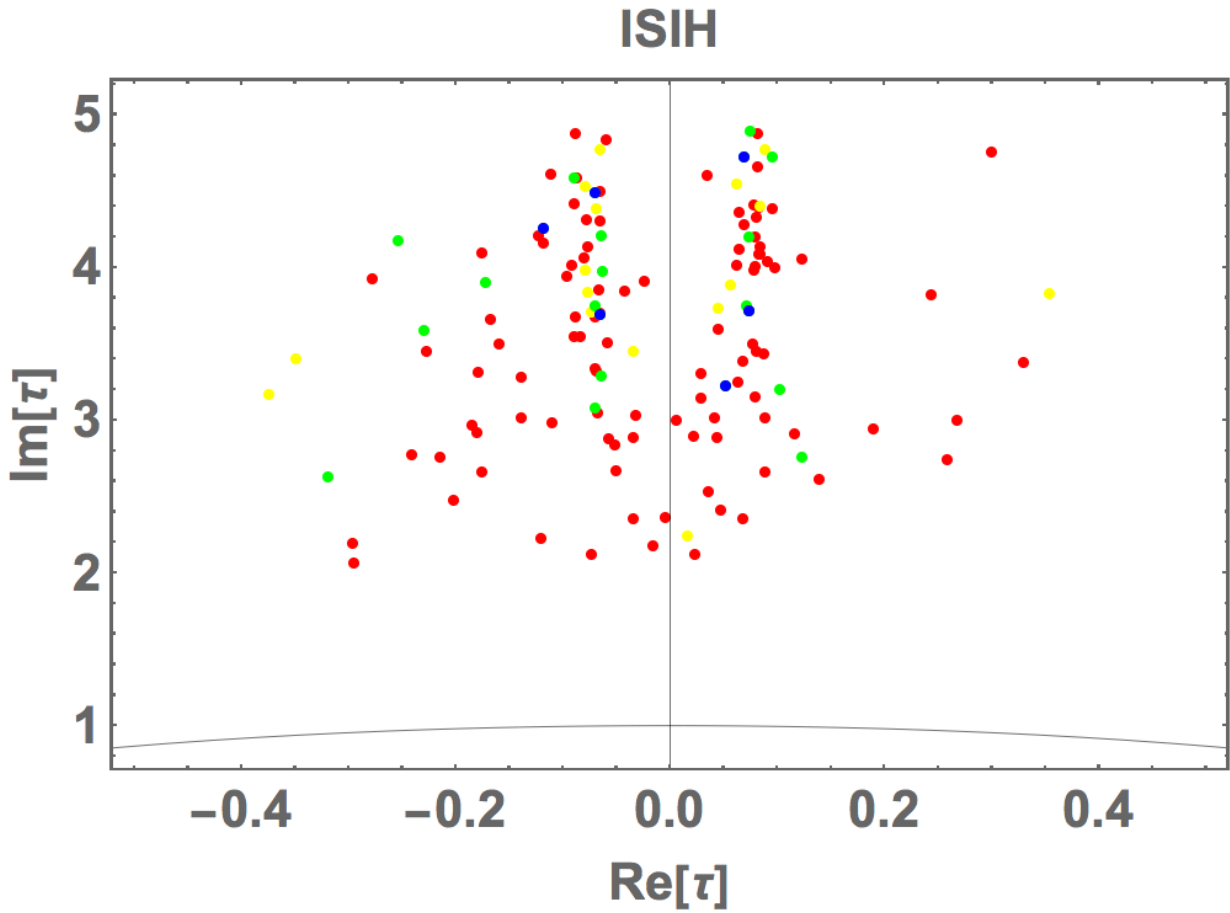}
\caption{Allowed region of real and imaginary part of $\tau$ in the fundamental region.
The blue, green, yellow, and red points respectively correspond to $\sigma\le1$, $1<\sigma\le2$, $2<\sigma\le3$ and  $3<\sigma\le5$ interval in $\Delta\chi^2$ analysis.. }
  \label{fig:tau_is}
\end{center}\end{figure}
In Fig.~\ref{fig:tau_is}, we plot the allowed region of real and imaginary part of $\tau$ in the fundamental region, where
the blue, green, yellow, and red points respectively correspond to $\sigma\le1$, $1<\sigma\le2$, $2<\sigma\le3$ and  $3<\sigma\le5$ interval in $\Delta\chi^2$ analysis.
In the NH case, the allowed region of real $\tau$ runs over the whole fundamental region while there is the upper bound on imaginary $\tau$ that is about 4. 
In the IH case, the allowed region of real $\tau$ runs over [$-0.4$,0.4] while the range of imaginary $\tau$ is [2,5].

\begin{figure}[tb]
\begin{center}
\includegraphics[width=77.0mm]{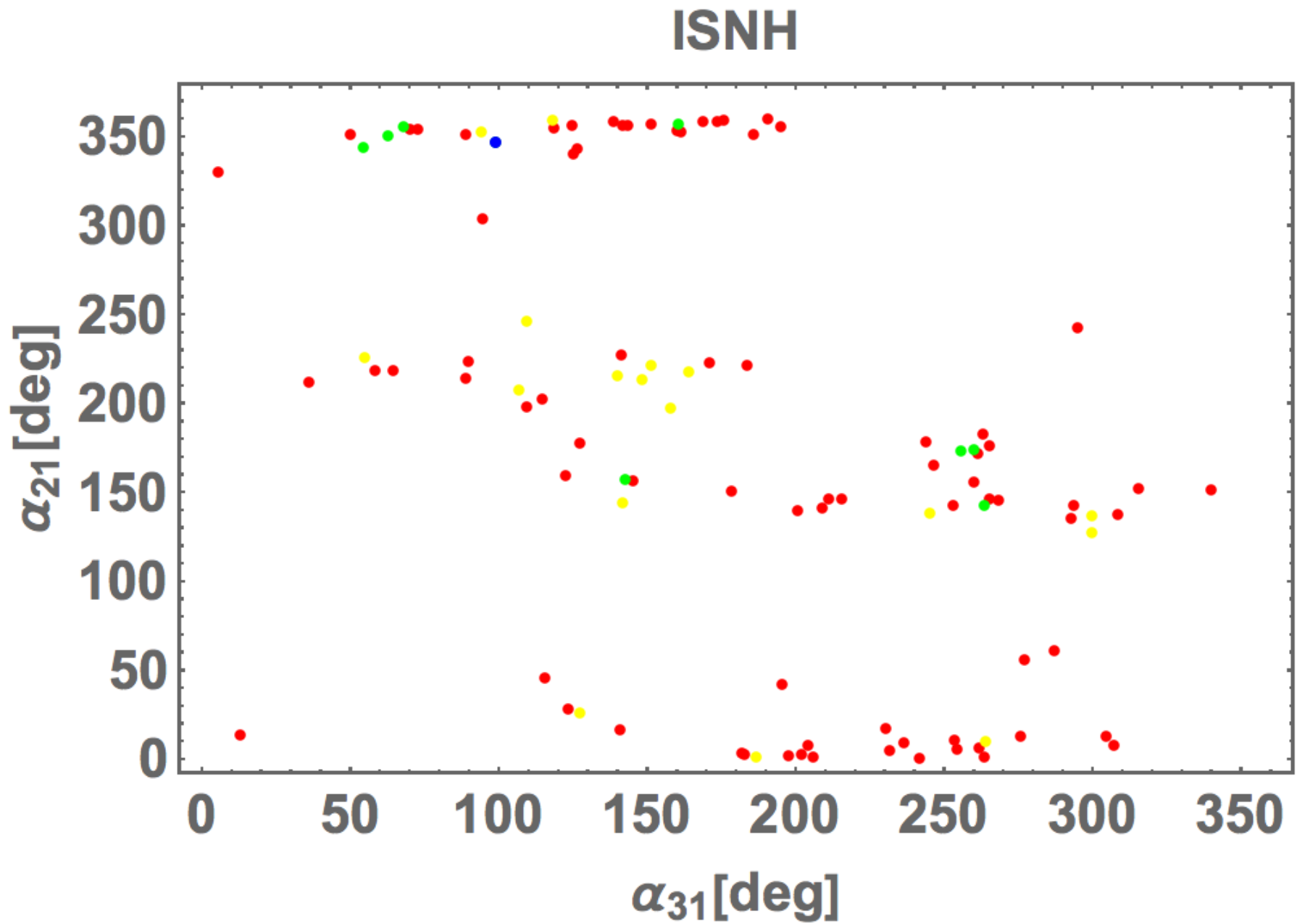}
\includegraphics[width=77.0mm]{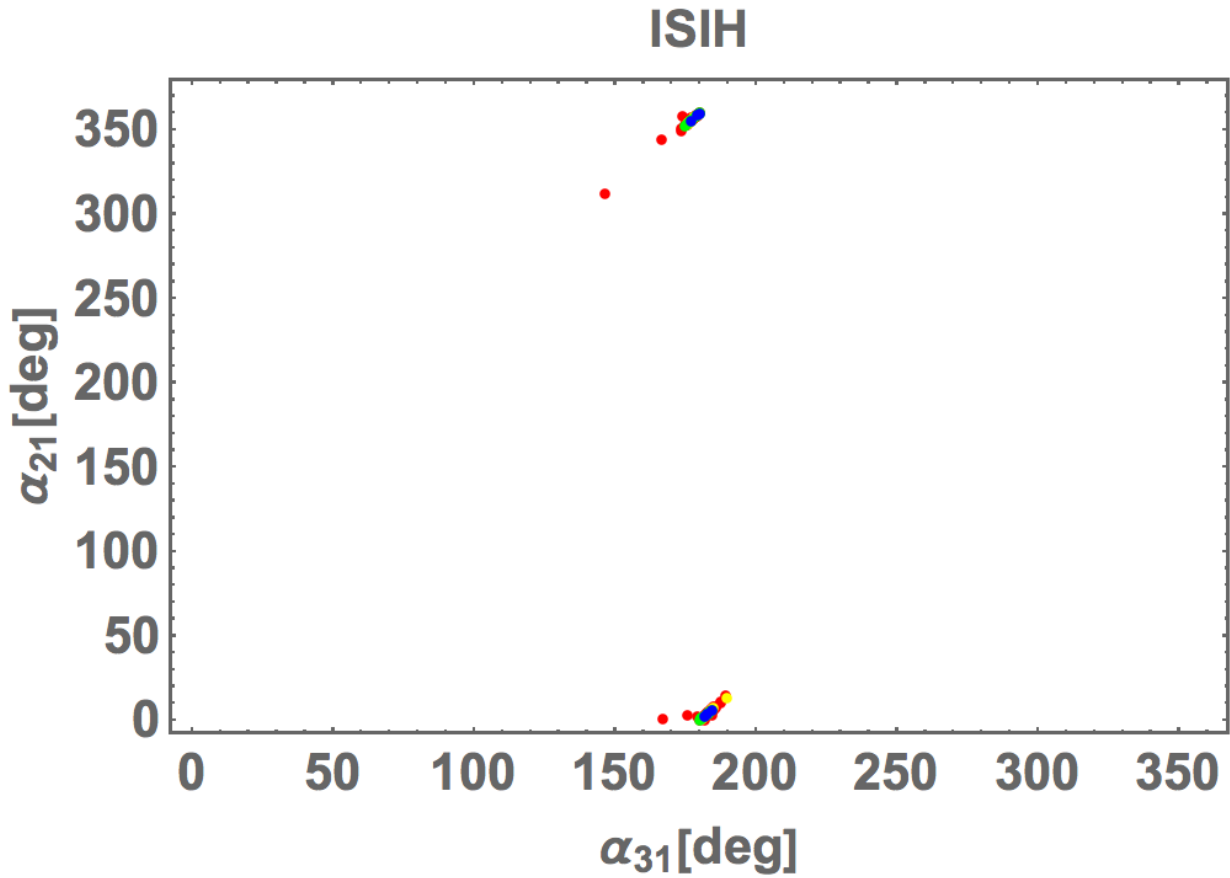}
\caption{Allowed region of Majorana phases, where the color legends are the same as the ones of Fig.~\ref{fig:tau_is}. }
  \label{fig:majo_is}
\end{center}\end{figure}
In Fig.~\ref{fig:majo_is}, we plot the allowed region of Majorana phases, where the color legends are the same as the ones of Fig.~\ref{fig:tau_is}.
 In case of NH, $\alpha_{31}$ runs the whole range but $\alpha_{21}$ is localized at nearby [$-50^\circ$,50$^\circ$]
and [130$^\circ$,250$^\circ$].
 In case of IH, $\alpha_{31}$ is [170$^\circ$,190$^\circ$] while $\alpha_{21}$ is [$-50^\circ$,20$^\circ$].

\begin{figure}[tb]
\begin{center}
\includegraphics[width=77.0mm]{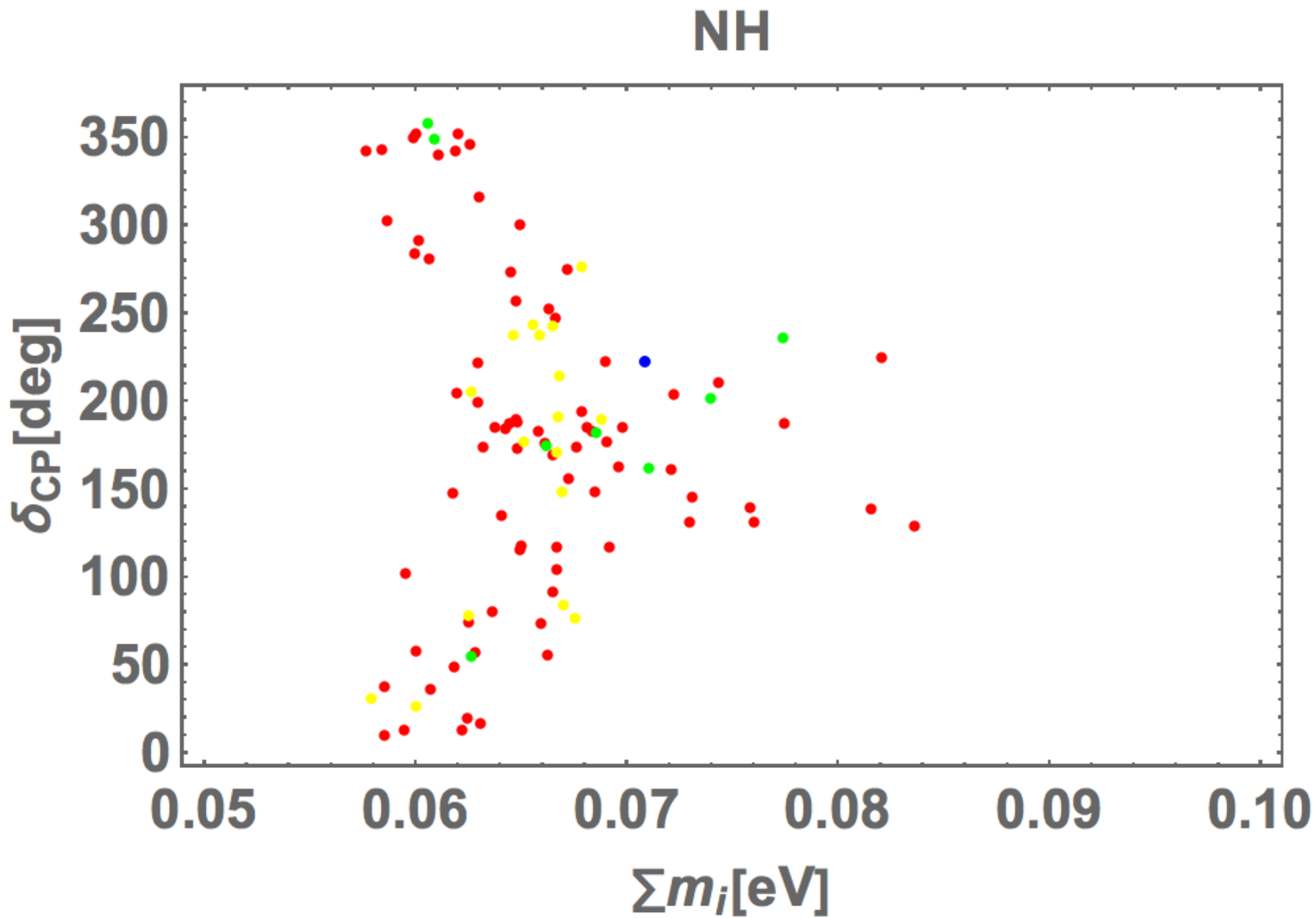}
\includegraphics[width=77.0mm]{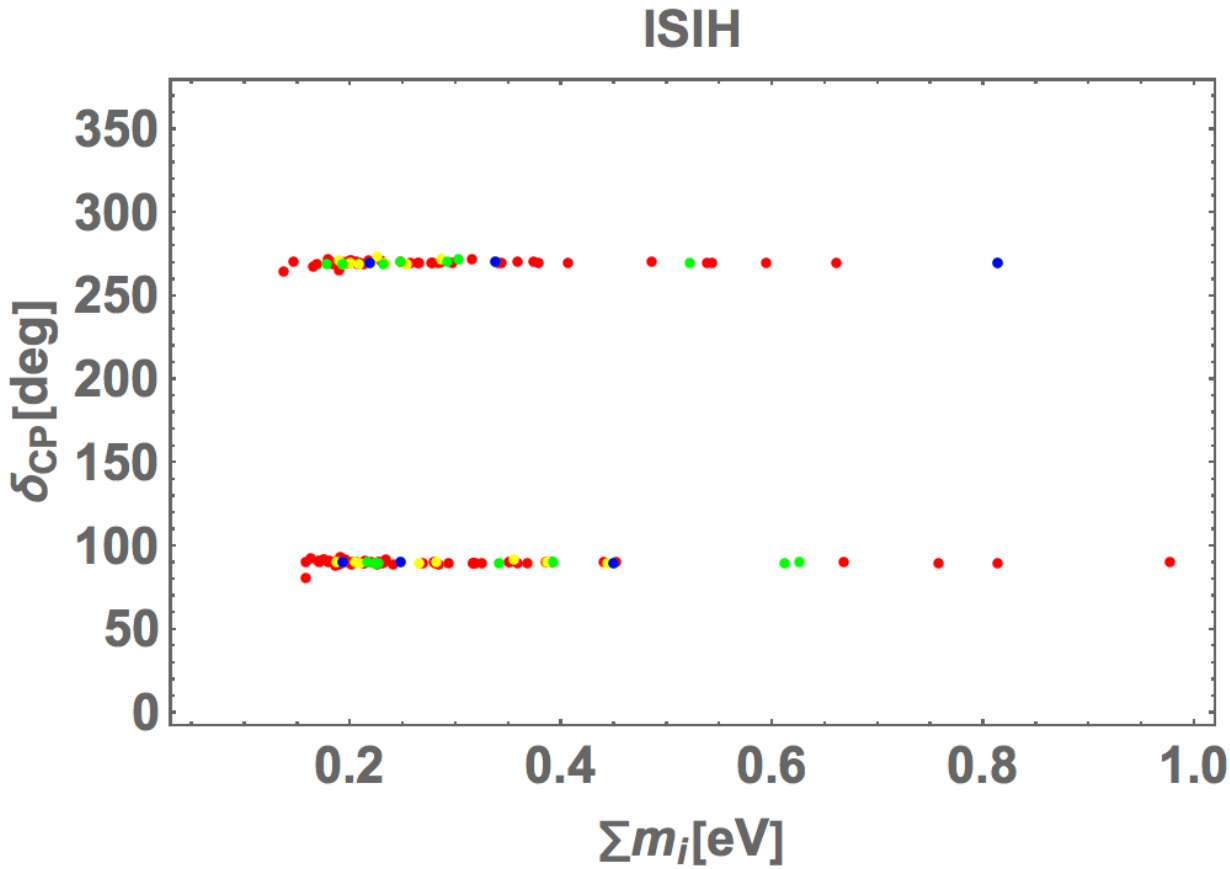}
\caption{Allowed region of sum of neutrino masses and CP Dirac phase, where the color legends are the same as the ones of Fig.~\ref{fig:tau_is}. }
  \label{fig:sumdcp_is}
\end{center}\end{figure}
In Fig.~\ref{fig:sumdcp_is}, we plot the allowed region of sum of neutrino masses and CP Dirac phase, where the color legends are the same as the ones of Fig.~\ref{fig:tau_is}.
 In case of NH, the sum of the active neutrino masses is allowed in the range of [0.059,0.085] eV while the whole range is allowed for $\delta_{CP}$.
 In case of IH, the sum of the active neutrino masses is allowed in the range of [0.14,1] eV and $\delta_{CP}$
 is localized by 90$^\circ$ and 270$^\circ$.

\begin{figure}[tb]
\begin{center}
\includegraphics[width=77.0mm]{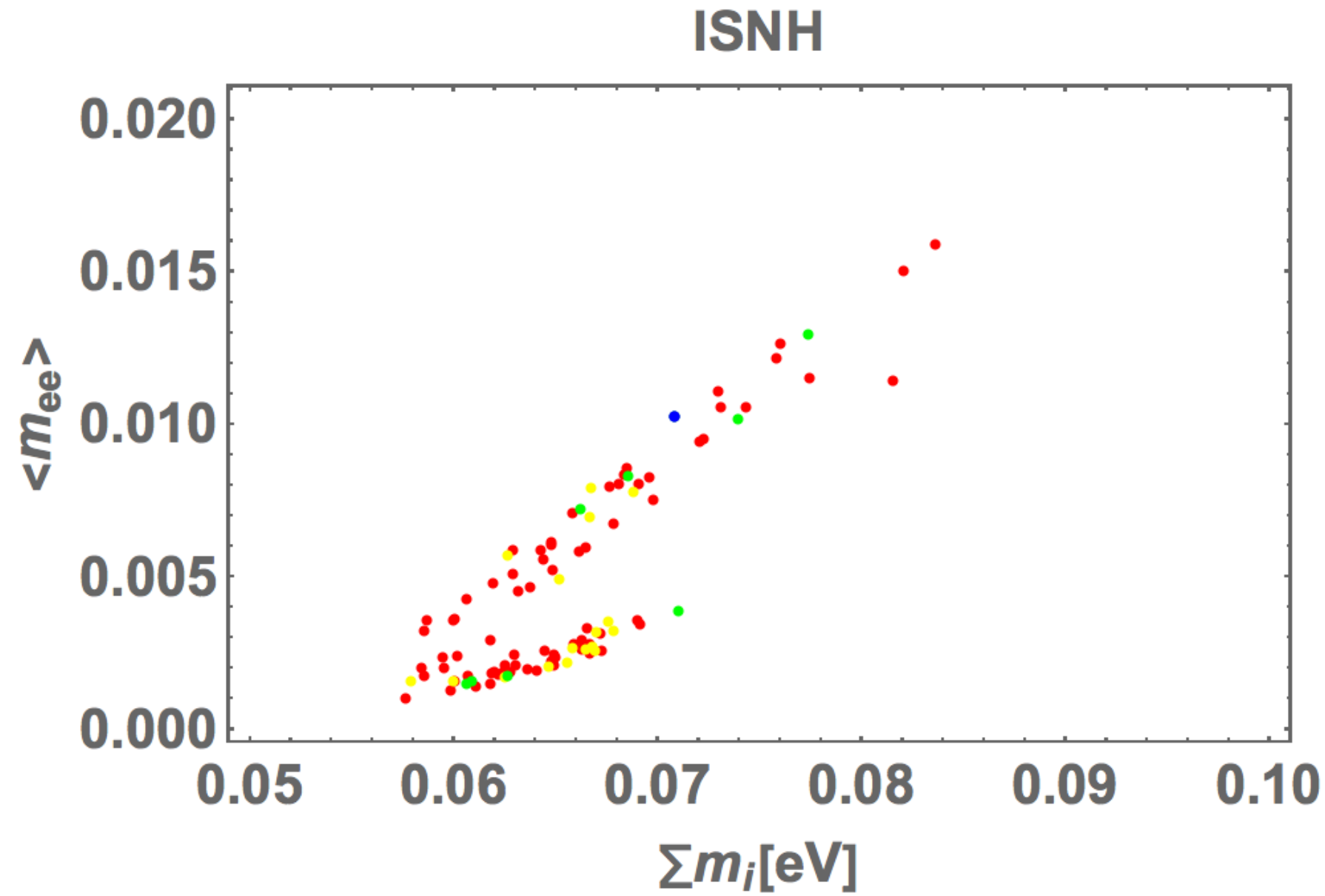}
\includegraphics[width=77.0mm]{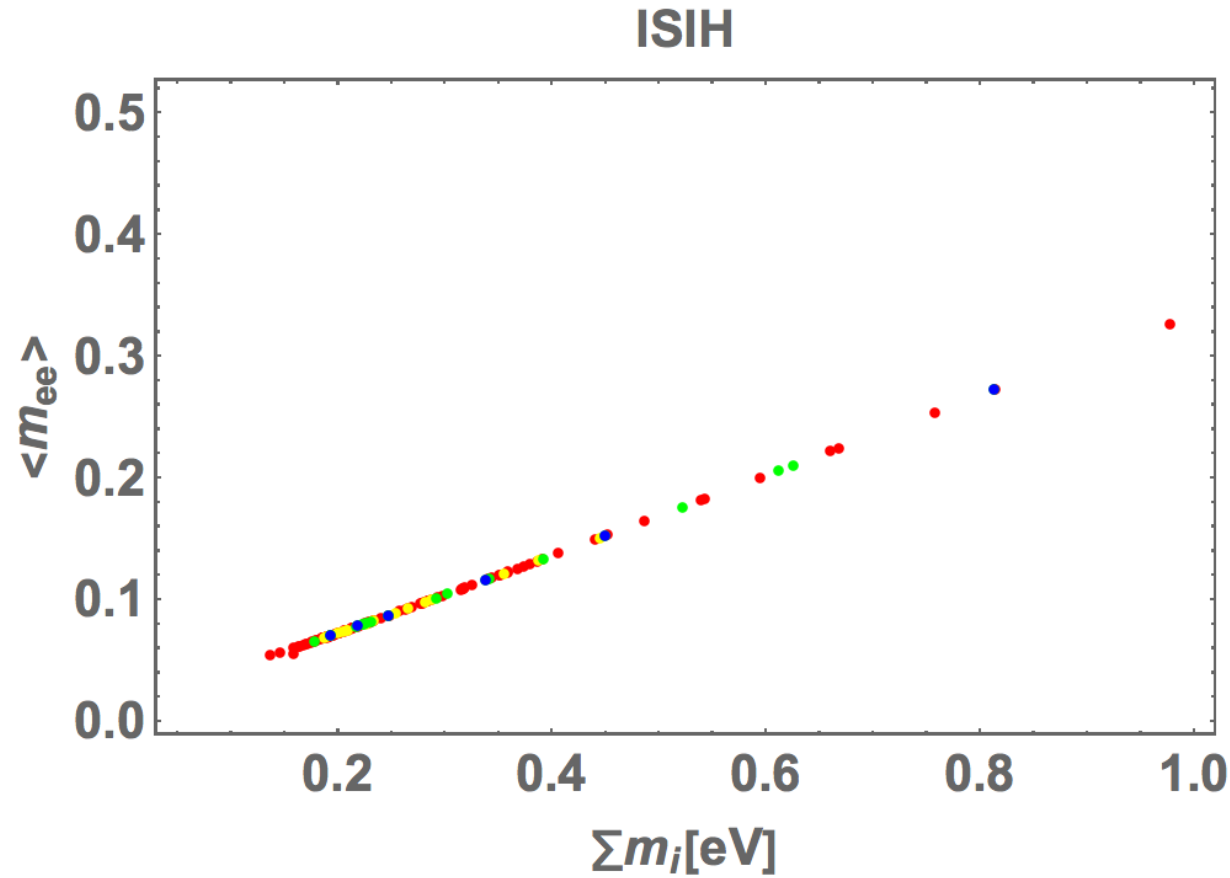}
\caption{Allowed region of sum of neutrino masses and the effective neutrinoless double beta decay, where the color legends are the same as the ones of Fig.~\ref{fig:tau_is}. }
  \label{fig:mases_is}
\end{center}\end{figure}
In Fig.~\ref{fig:mases_is}, we plot the allowed region of sum of neutrino masses and the effective neutrinoless double beta decay, where the color legends are the same as the ones of Fig.~\ref{fig:tau_is}.
 In case of NH, the effective neutrinoless double beta decay is allowed in the range of [0.001,0.016] eV.
 In case of IH, the effective neutrinoless double beta decay is allowed in the range of [0.05,0.3] eV.

\begin{figure}[tb]
\begin{center}
\includegraphics[width=77.0mm]{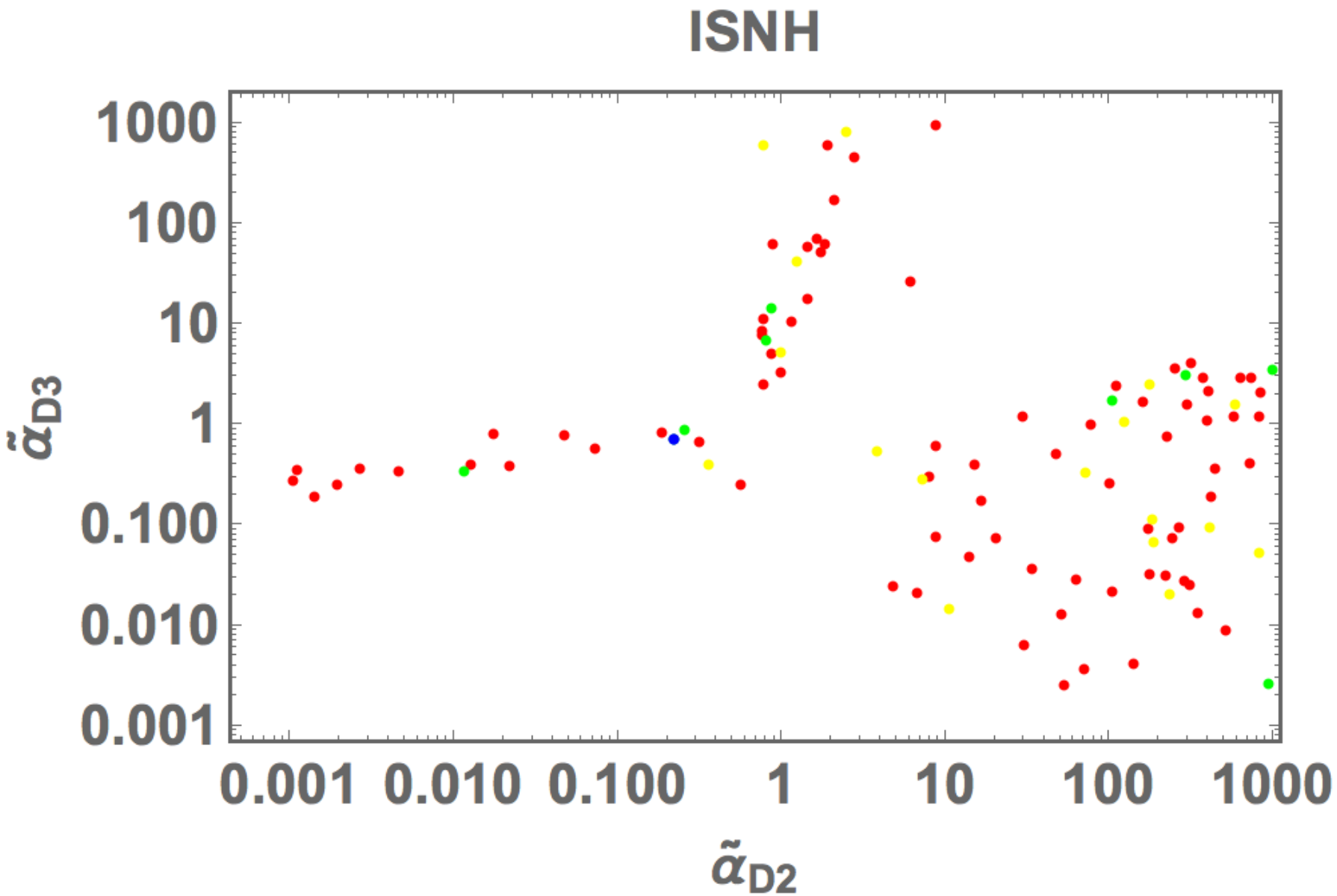}
\includegraphics[width=77.0mm]{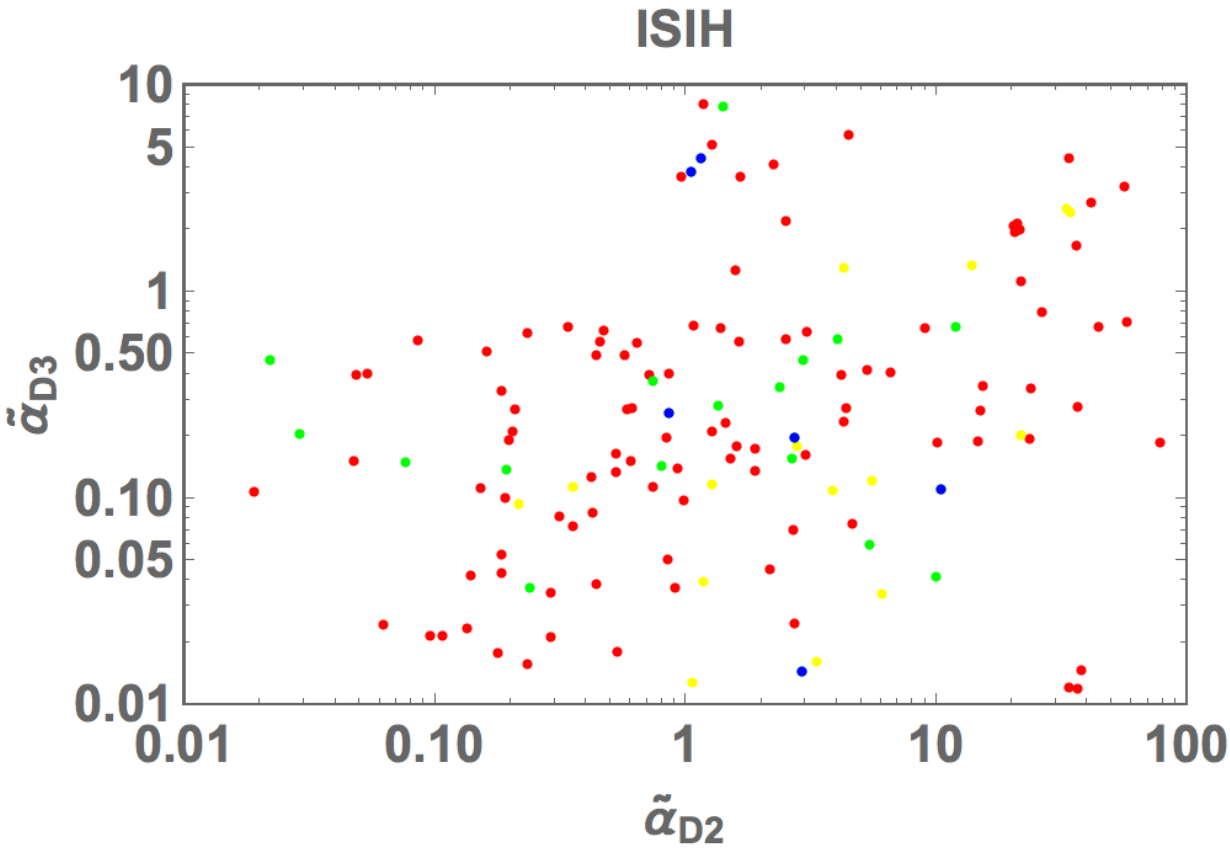}
\caption{Allowed region of input parameters of $\tilde\alpha_{2,3}$, where the color legends are the same as the ones of Fig.~\ref{fig:tau_is}. }
  \label{fig:alpha_is}
\end{center}\end{figure}
In Fig.~\ref{fig:alpha_is}, we plot the allowed region of $\tilde\alpha_{2,3}$, where the color legends are the same as the ones of Fig.~\ref{fig:tau_is}.
 In case of NH, although $\tilde\alpha_{2,3}$ runs the whole range which we set, there is a correlation between them.
 In case of IH,  $\tilde\alpha_{2,3}$ runs $[0.01,100]$ but their correlation is not so stronger than the case of NH.

\begin{figure}[tb]
\begin{center}
\includegraphics[width=77.0mm]{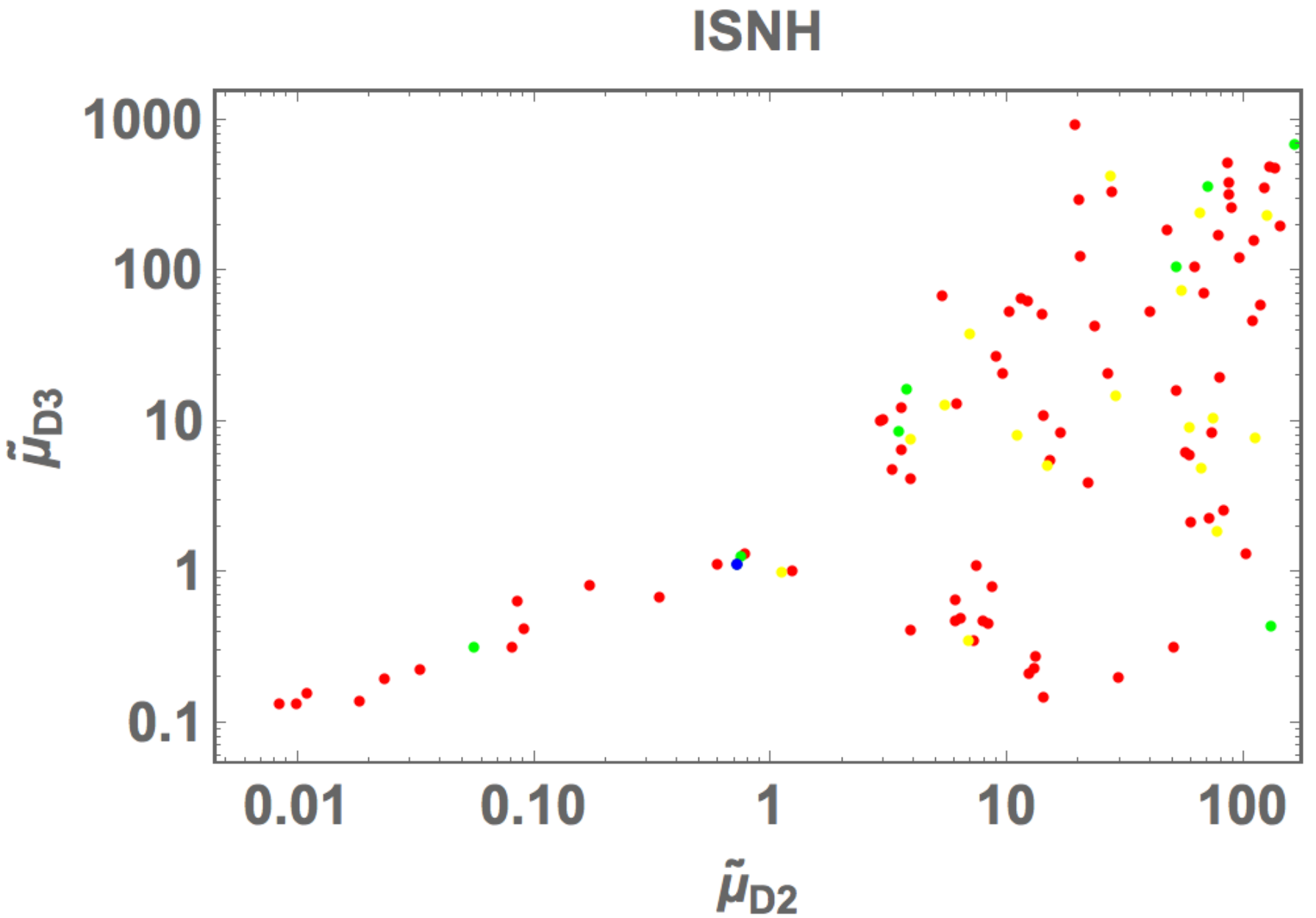}
\includegraphics[width=77.0mm]{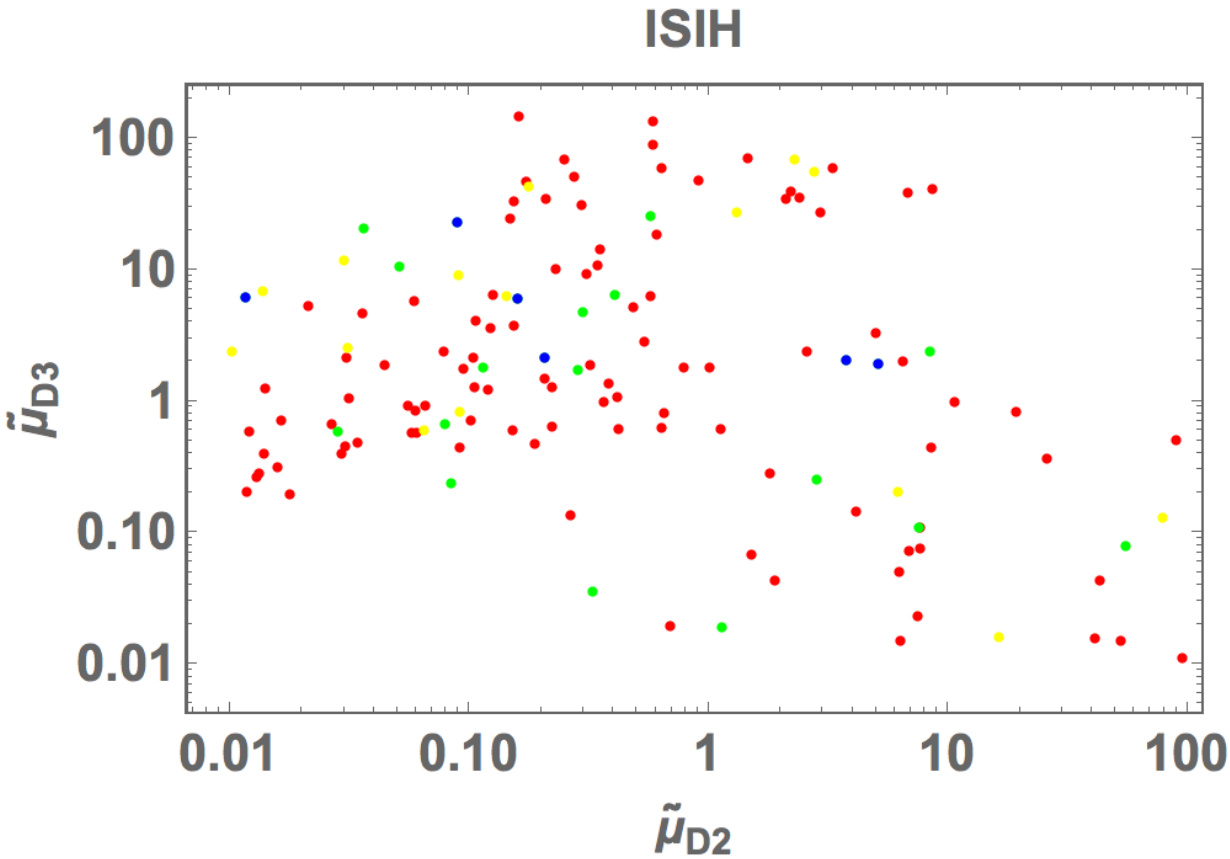}
\caption{Allowed region of input parameters of $\tilde\mu_{2,3}$, where the color legends are the same as the ones of Fig.~\ref{fig:tau_is}. }
  \label{fig:mu_is}
\end{center}\end{figure}
In Fig.~\ref{fig:mu_is}, we plot the allowed region of $\tilde\mu_{2,3}$, where the color legends are the same as the ones of Fig.~\ref{fig:tau_is}.
 In case of NH, $\tilde\mu_{2}$ runs [0.01,100] and $\tilde\mu_{3}$ runs [0.1,1000]. Furthermore, there is a correlation between them.
 In case of IH,  $\tilde\mu_{2,3}$ runs $[0.01,100]$ but their correlation is not so stronger than the case of NH.

\subsubsection{LS}
We randomly select our free parameters within the following range:
\begin{align}
\{\tilde \alpha_{D_{2,3}},\tilde \alpha_{D'_{2,3}}, |\tilde \mu_{NS_{2,3}}|\}\in [10^{-3} -10^3],
\end{align}
and we work on the fundamental region on $\tau$. 

\begin{figure}[tb]
\begin{center}
\includegraphics[width=77.0mm]{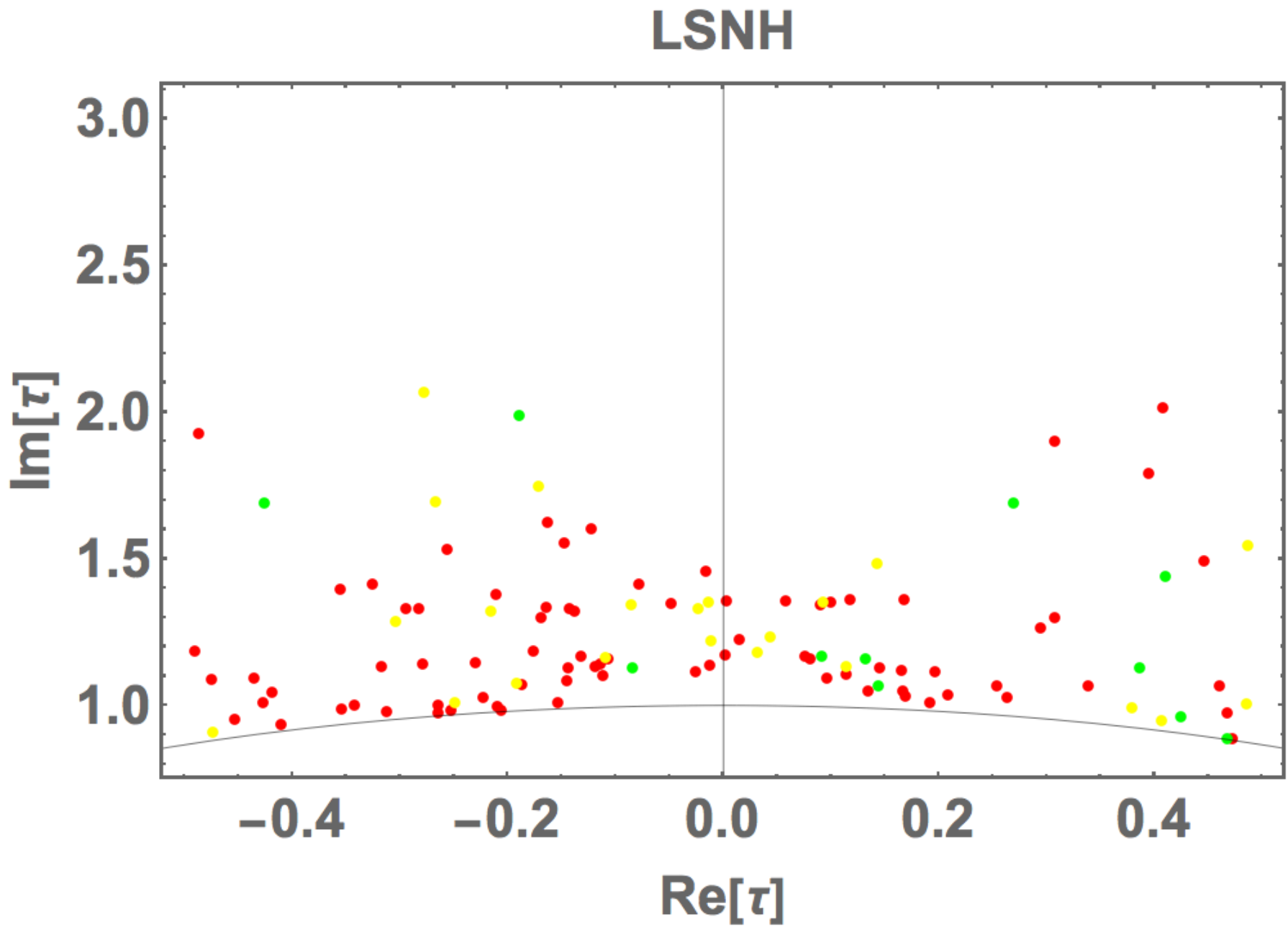}
\includegraphics[width=77.0mm]{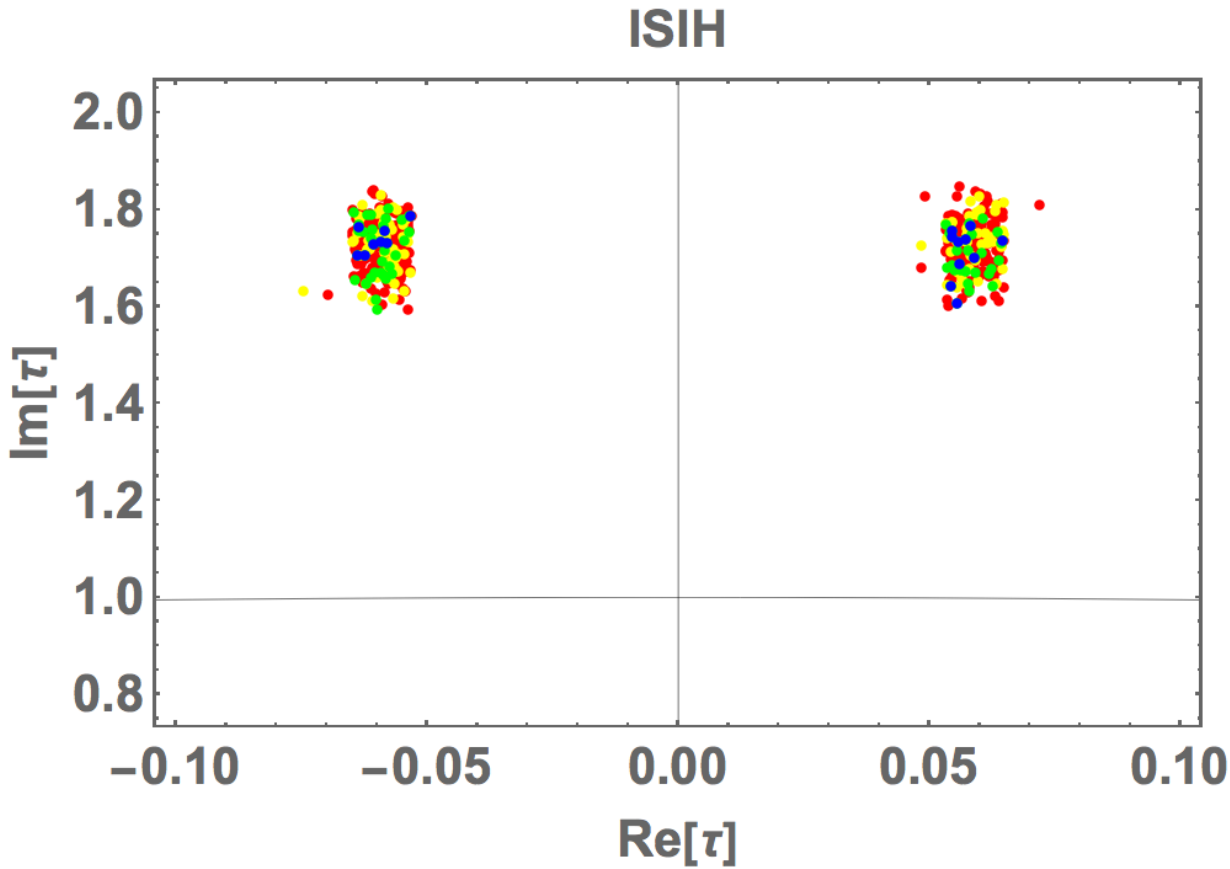}
\caption{Allowed region of real and imaginary part of $\tau$ in the fundamental region where the color legends are the same as the ones of Fig.~\ref{fig:tau_is}.  }
  \label{fig:tau_ls}
\end{center}\end{figure}
In Fig.~\ref{fig:tau_ls}, we plot the allowed region of real and imaginary part of $\tau$ in the fundamental region where the color legends are the same as the ones of Fig.~\ref{fig:tau_is}.
In the case NH, real $\tau$ runs the whole range but imaginary $\tau$ has the upper bound Im$[\tau]\le2$.
In the case IH,
$\tau$ is localized at nearby 
$|{\rm Re}[\tau]|=[0.05,0.07]$ and Im$[\tau]=[1.6,1.8]$.

\begin{figure}[tb]
\begin{center}
\includegraphics[width=77.0mm]{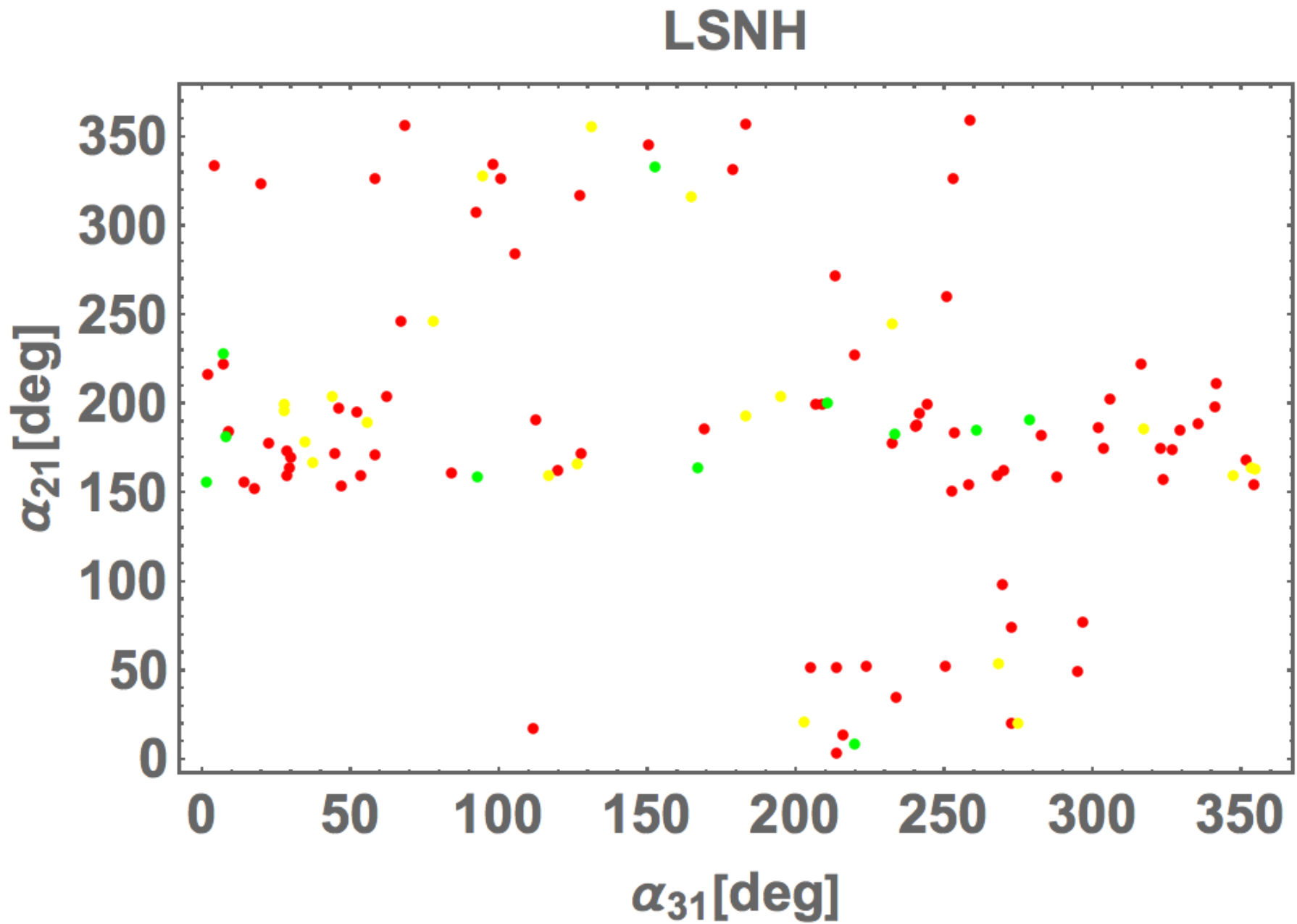}
\includegraphics[width=77.0mm]{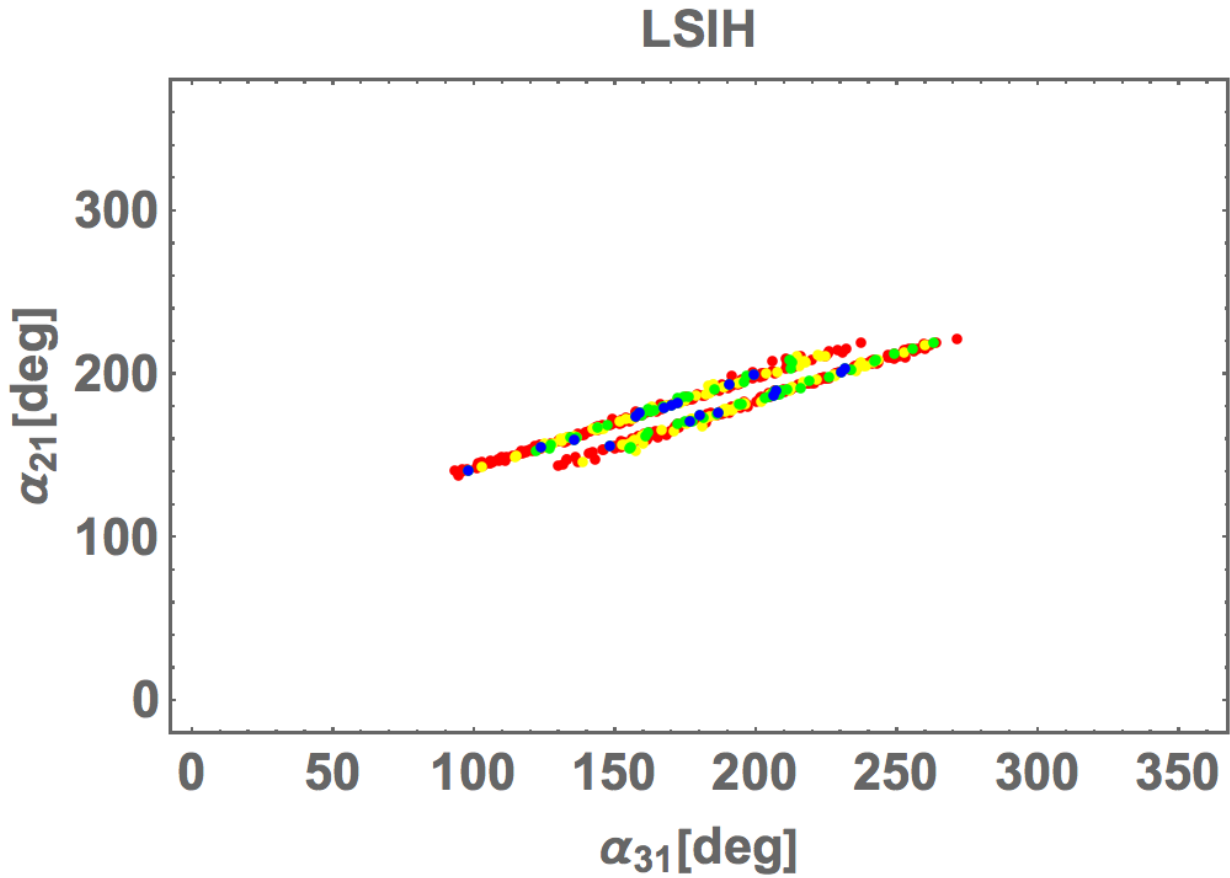}
\caption{Allowed region of Majorana phases, where the color legends are the same as the ones of Fig.~\ref{fig:tau_is}. }
  \label{fig:majo_ls}
\end{center}\end{figure}
In Fig.~\ref{fig:majo_ls}, we plot the allowed region of Majorana phases, where the color legends are the same as the ones of Fig.~\ref{fig:tau_is}.
In the case NH, even though the whole ranges are allowed, $\alpha_{21}$ tends to be localized at nearby 180$^\circ$.
In the case IH, $\alpha_{31}$ is localized at nearby $[90^\circ,270^\circ]$ and $\alpha_{21}$ is also localized at nearby $[120^\circ,220^\circ]$.

\begin{figure}[tb]
\begin{center}
\includegraphics[width=77.0mm]{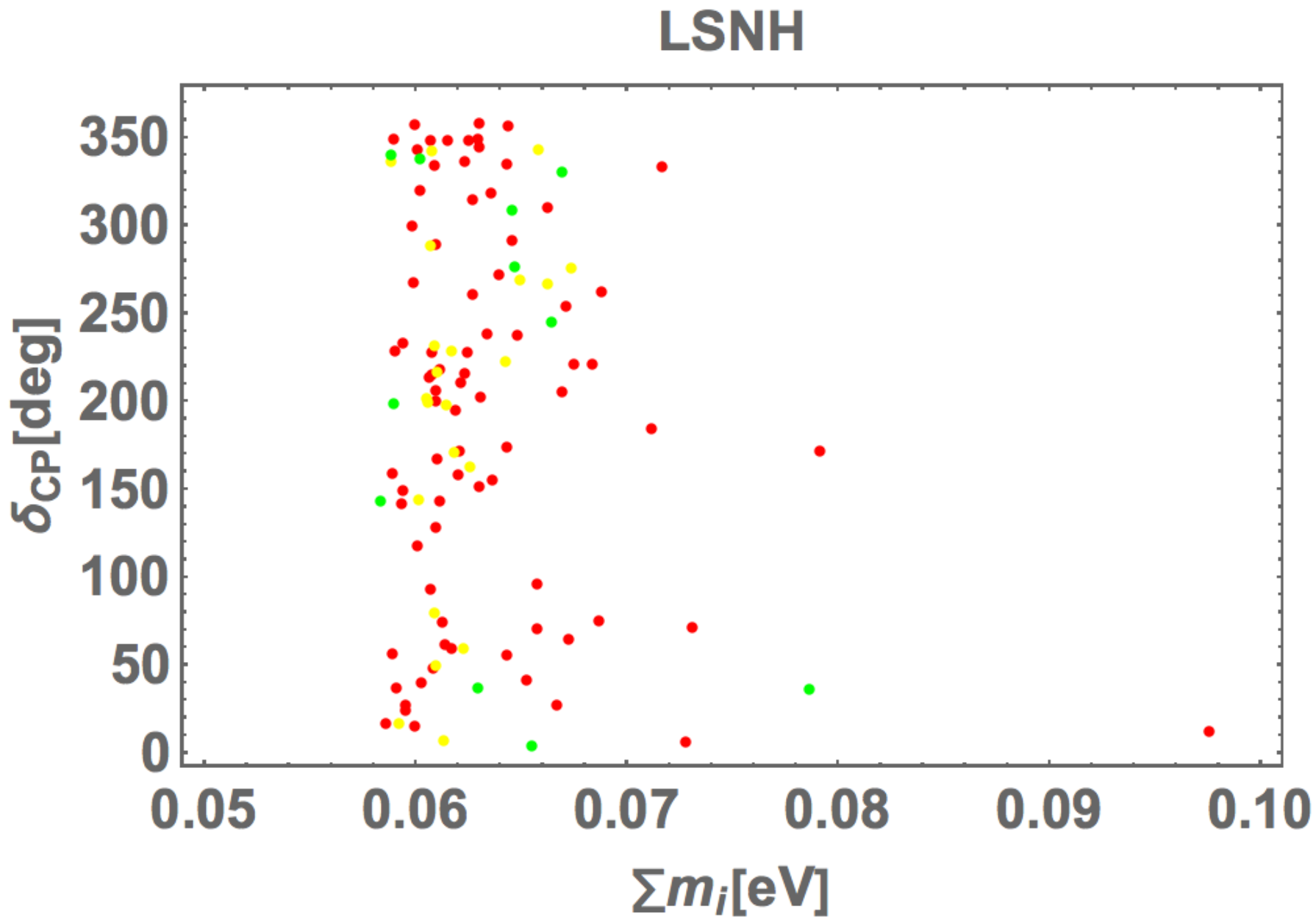}
\includegraphics[width=77.0mm]{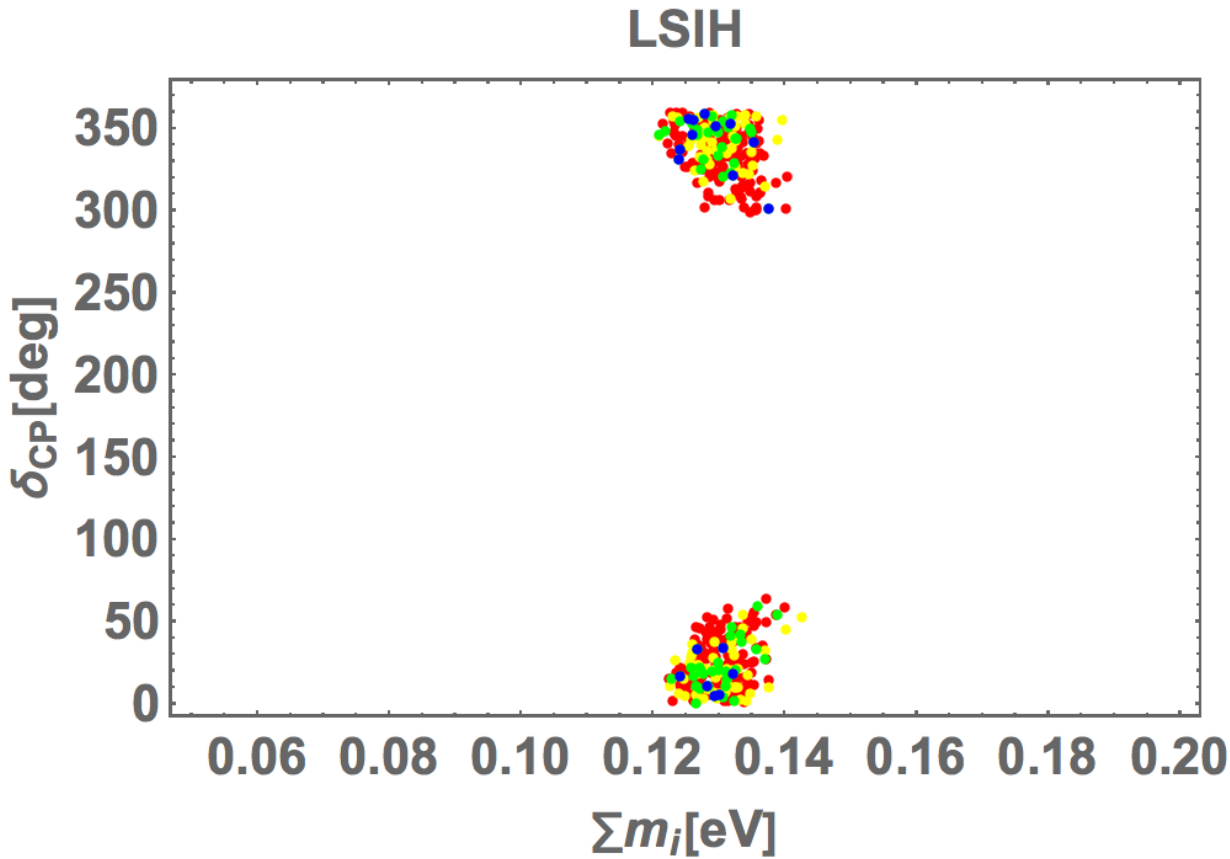}
\caption{Allowed region of sum of neutrino masses and CP Dirac phase, where the color legends are the same as the ones of Fig.~\ref{fig:tau_is}. }
  \label{fig:sumdcp_ls}
\end{center}\end{figure}
In Fig.~\ref{fig:sumdcp_ls}, we plot the allowed region of sum of neutrino masses and CP Dirac phase, where the color legends are the same as the ones of Fig.~\ref{fig:tau_is}.
In the NH case, the whole range is allowed for $\delta_{CP}$ and sum of the active neutrino masses runs over the range of [0.059,0.098] eV.
In the IH case, $\sum m_i$ is localized at nearby $[0.12,0.14]$ eV, and $\delta_{CP}$ is localized at nearby $[-60^\circ,60^\circ]$.

\begin{figure}[tb]
\begin{center}
\includegraphics[width=77.0mm]{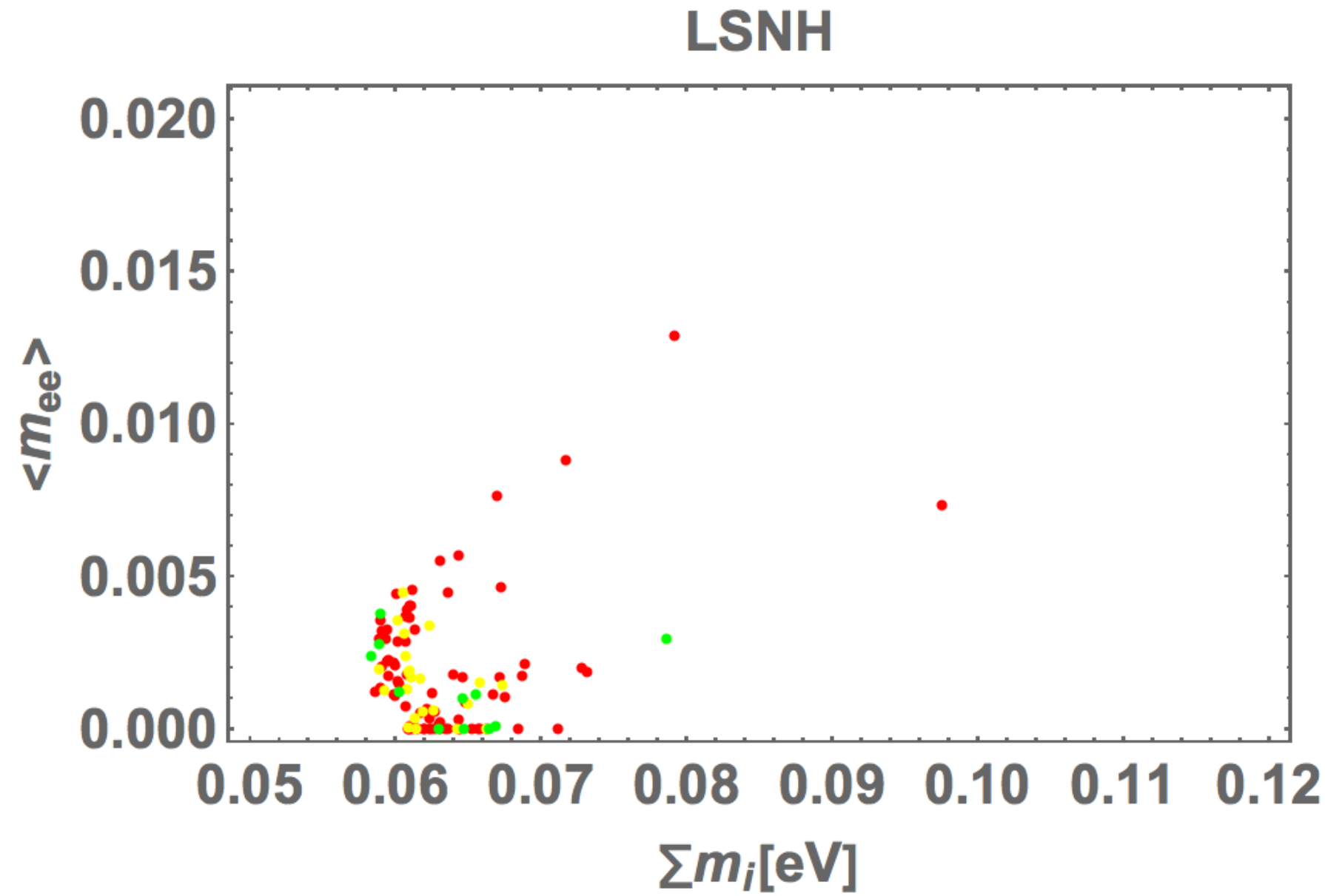}
\includegraphics[width=77.0mm]{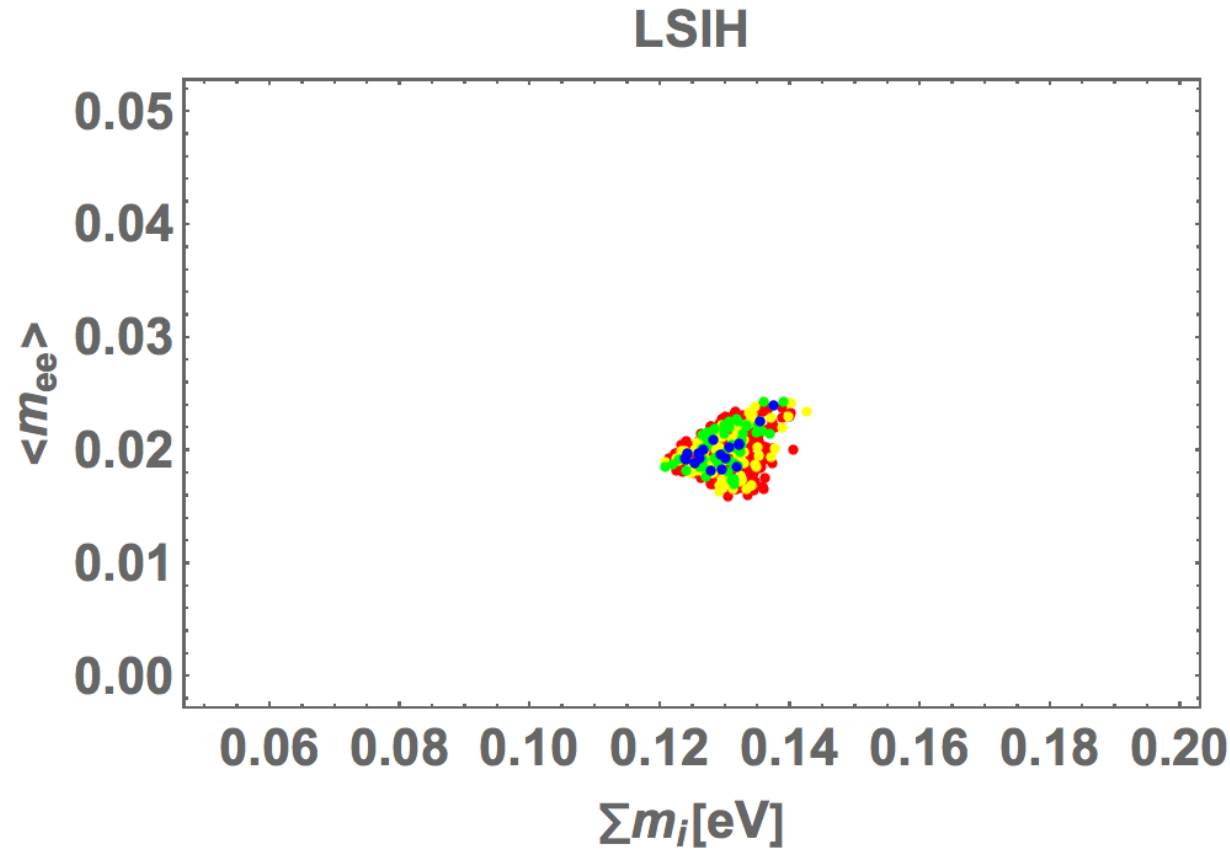}
\caption{Allowed region of sum of neutrino masses and the effective neutrinoless double beta decay, where the color legends are the same as the ones of Fig.~\ref{fig:tau_is}. }
  \label{fig:mases_ls}
\end{center}\end{figure}
In Fig.~\ref{fig:mases_ls}, we plot the allowed region of sum of neutrino masses and the effective neutrinoless double beta decay, where the color legends are the same as the ones of Fig.~\ref{fig:tau_is}.
In the NH case, $\langle m_{ee}\rangle$ runs over [0,0.013] eV.
In the IH case, $\langle m_{ee}\rangle$ is localized at nearby [0.016,0.024] eV.

\begin{figure}[tb]
\begin{center}
\includegraphics[width=77.0mm]{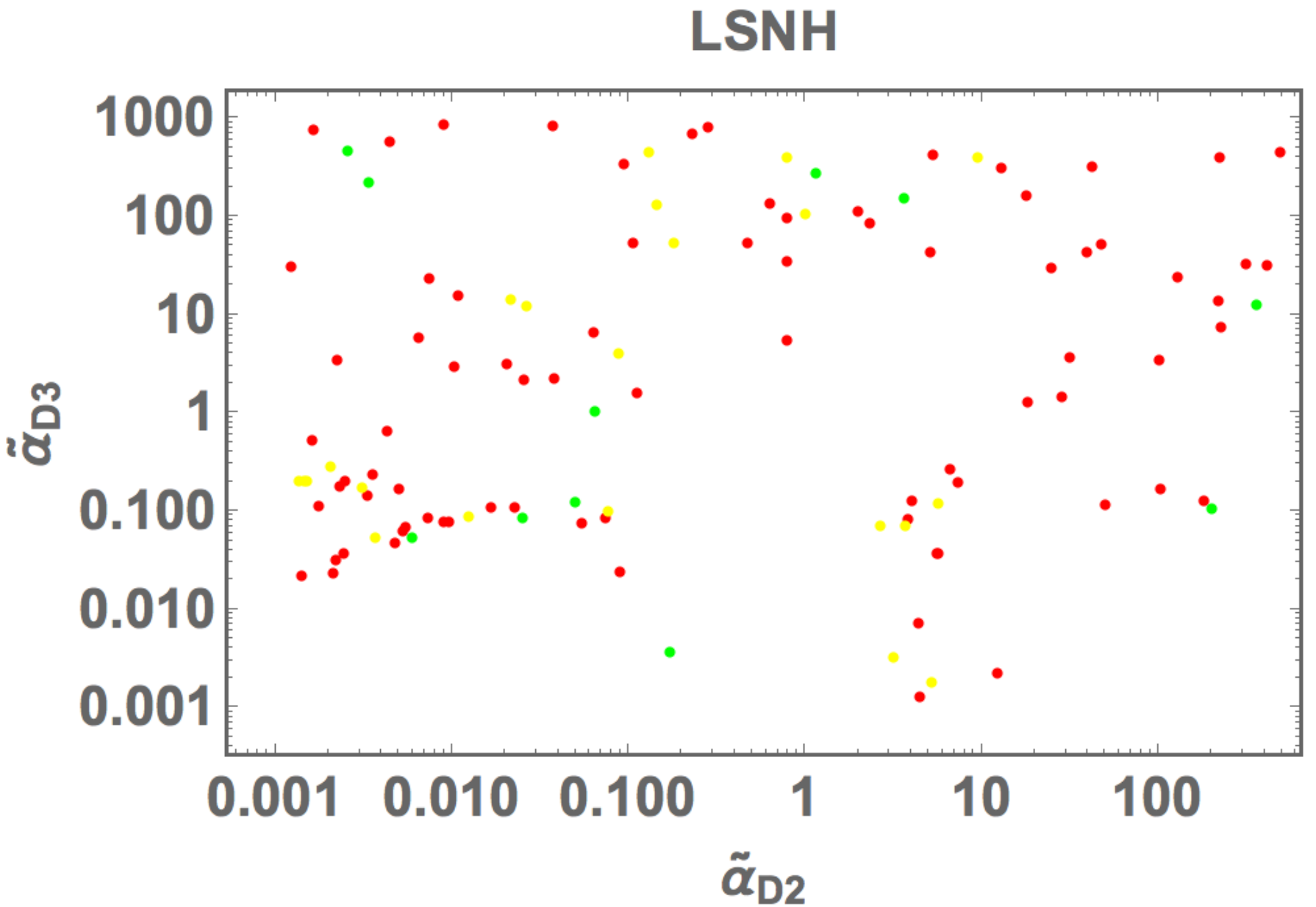}
\includegraphics[width=77.0mm]{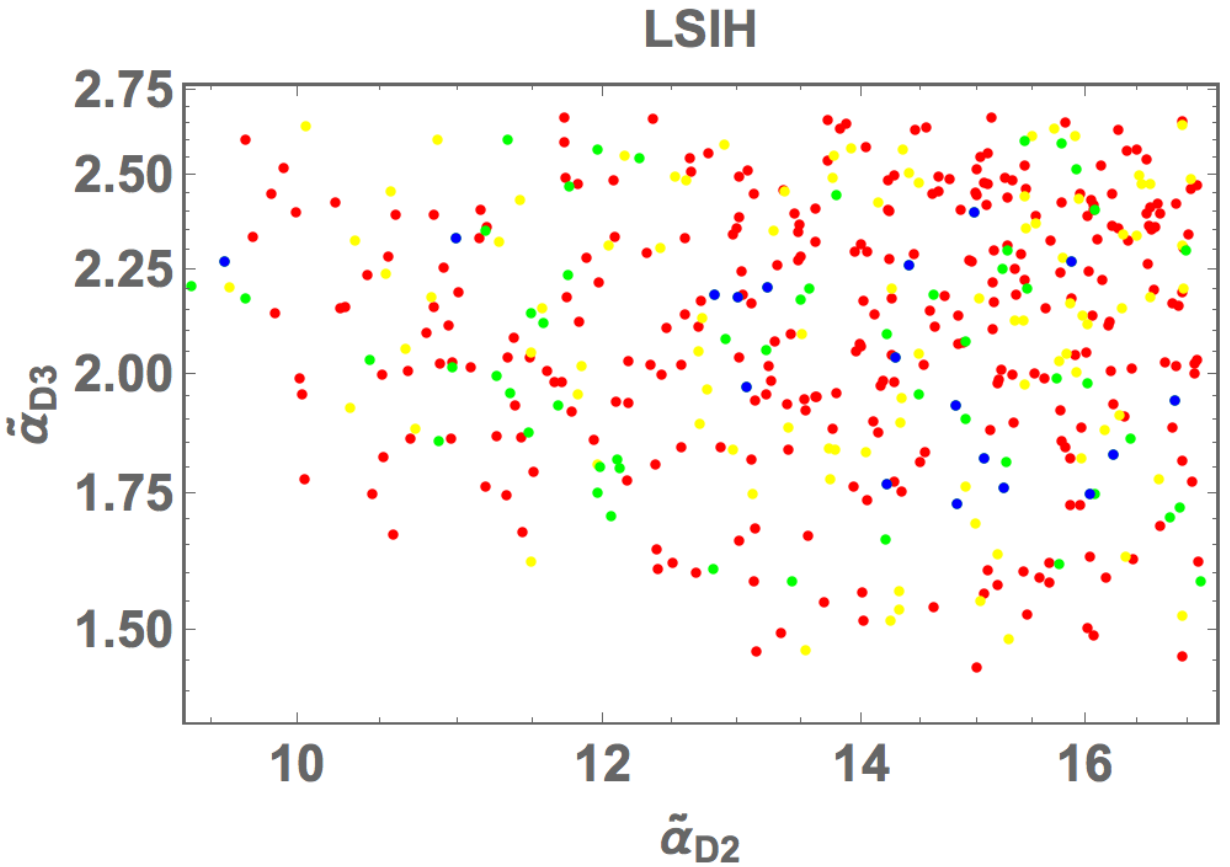}
\caption{Allowed region of input parameters of $\tilde\alpha_{D_{2,3}}$, where the color legends are the same as the ones of Fig.~\ref{fig:tau_is}. }
  \label{fig:alpha_ls}
\end{center}\end{figure}
In Fig.~\ref{fig:alpha_ls}(\ref{fig:alpha2_ls}), we plot the allowed region of $\tilde\alpha_{D_{2,3}}(\tilde\alpha_{D'_{2,3}})$, where the color legends are the same as the ones of Fig.~\ref{fig:tau_is}. In the NH case, $\tilde\alpha_{D_{2,3}}(\tilde\alpha_{D'_{2,3}})$ runs the whole ranges and their correlations are weak.
In the IH case, $\tilde\alpha_{D_{2}}$ and $\tilde\alpha_{D_{3}}$ are respectively localized at nearby [9,17] and [1.49,2.75].
$\tilde\alpha_{D'_{2}}$ and $\tilde\alpha_{D'_{3}}$ are respectively localized at nearby [0.006,0.012] and [200,400].

\begin{figure}[tb]
\begin{center}
\includegraphics[width=77.0mm]{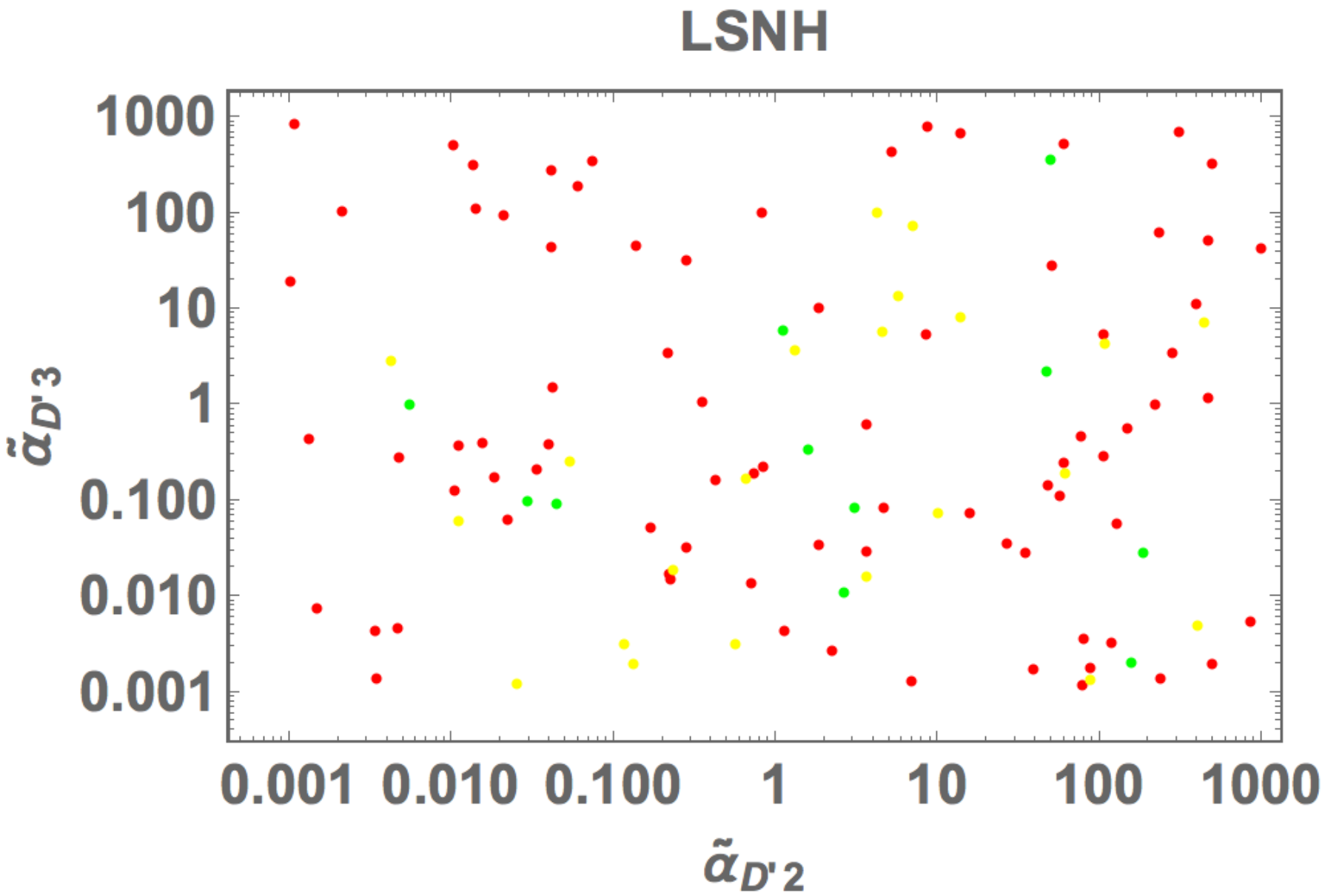}
\includegraphics[width=77.0mm]{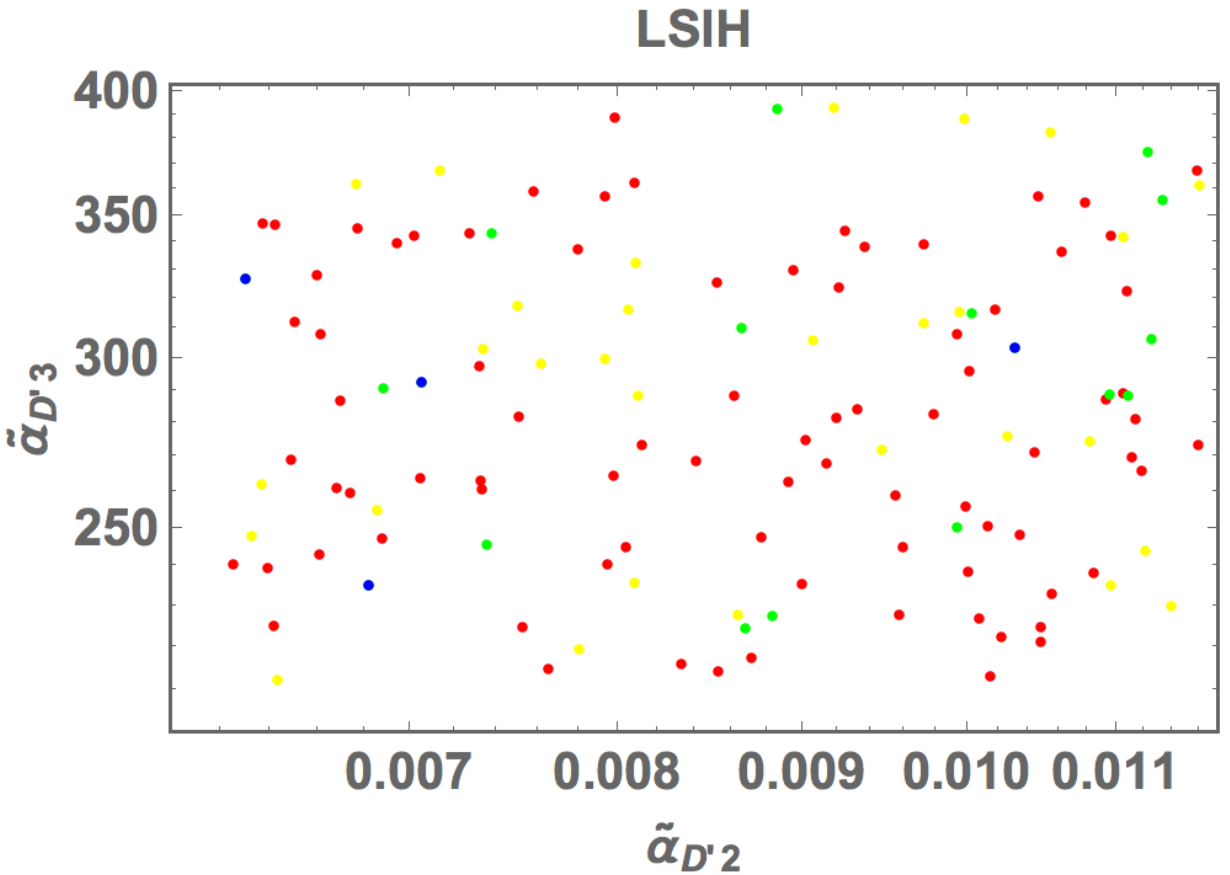}
\caption{Allowed region of input parameters of $\tilde\alpha_{D'_{2,3}}$, where the color legends are the same as the ones of Fig.~\ref{fig:tau_is}. }
  \label{fig:alpha2_ls}
\end{center}\end{figure}

\begin{figure}[tb]
\begin{center}
\includegraphics[width=77.0mm]{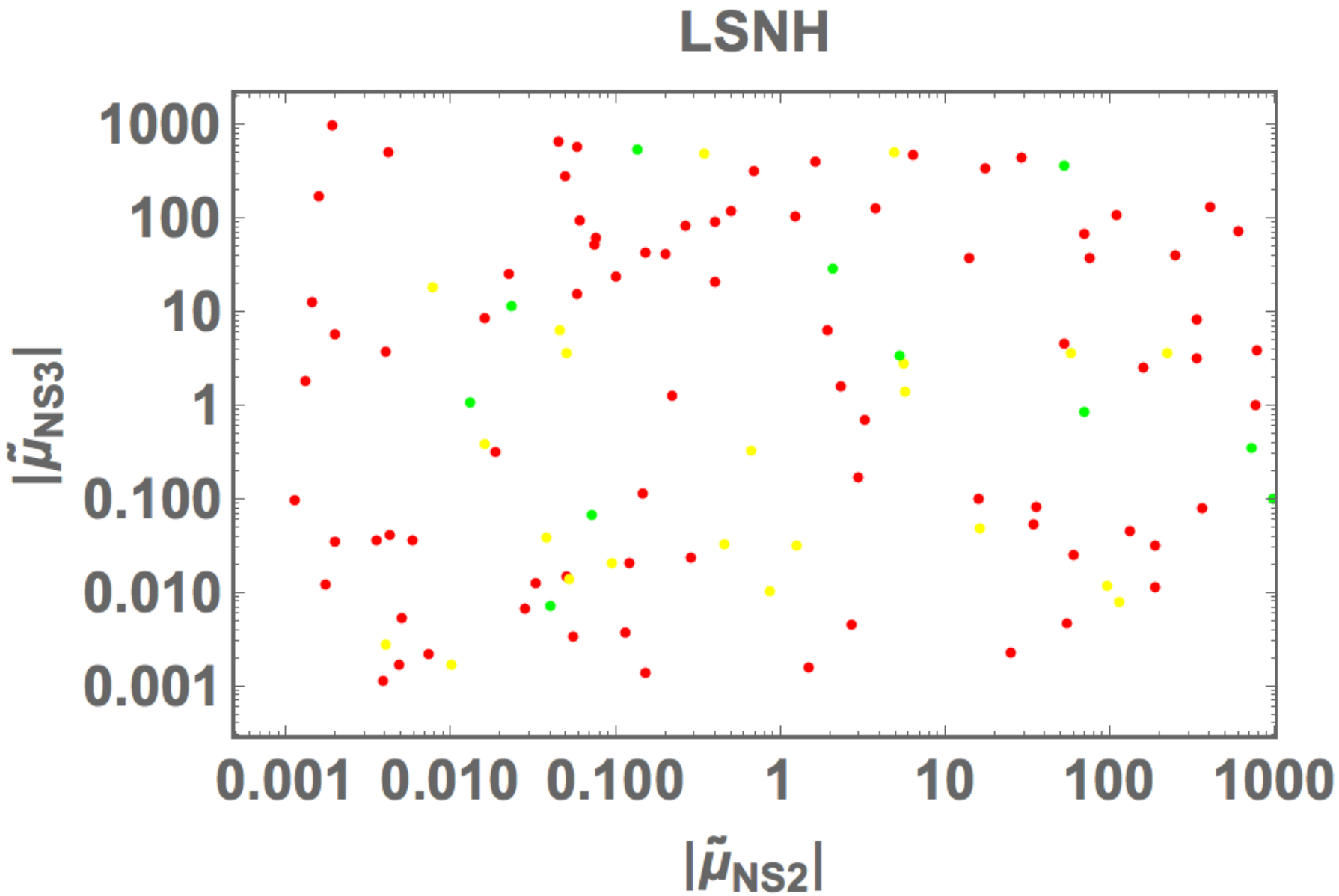}
\includegraphics[width=77.0mm]{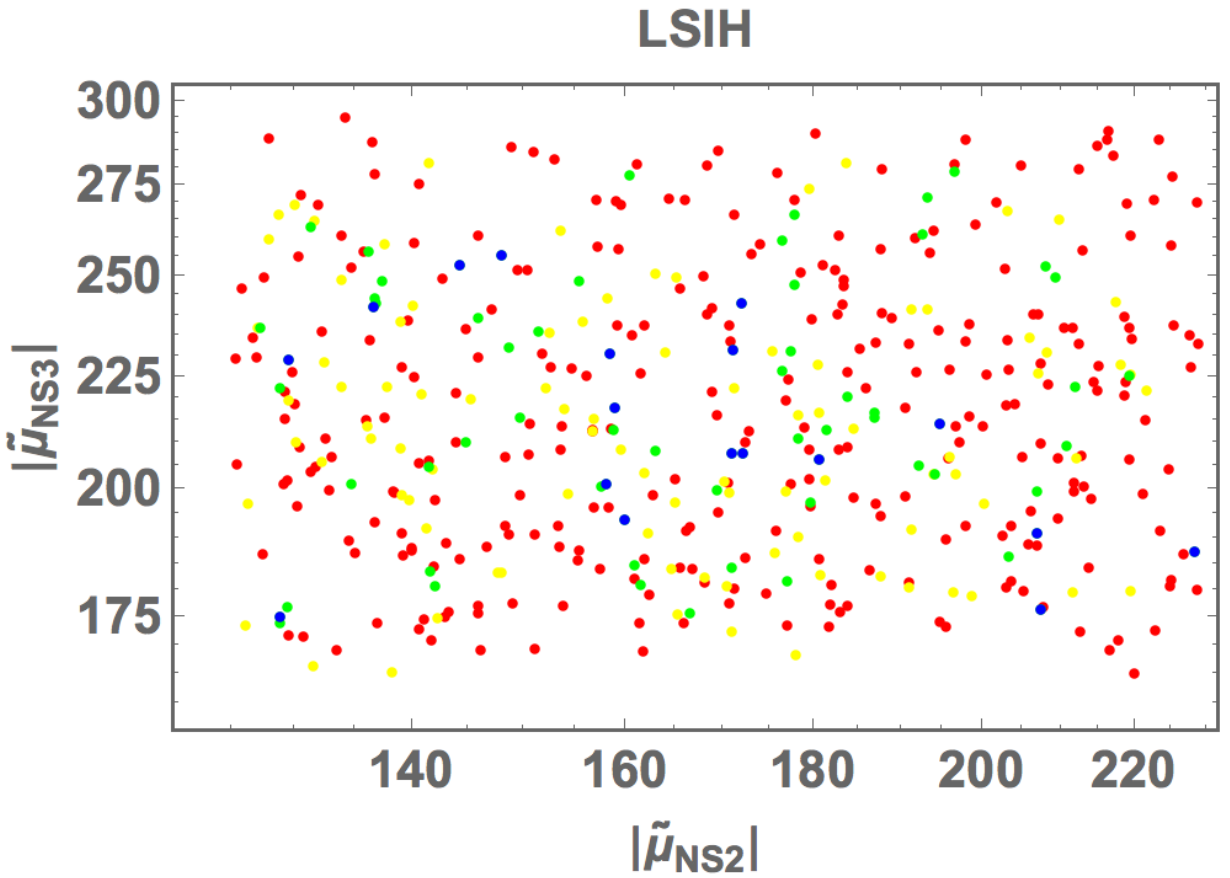}
\caption{Allowed region of input parameters of $\tilde\mu_{NS_{2,3}}$, where the color legends are the same as the ones of Fig.~\ref{fig:tau_is}. The left figure is their absolute values while the right one is their arguments.}
  \label{fig:mu_ls}
\end{center}\end{figure}
In Fig.~\ref{fig:mu_ls}, we plot the allowed region of $\tilde\mu_{NS_{2,3}}$, where the color legends are the same as the ones of Fig.~\ref{fig:tau_is}.
The left figure is their absolute values while the right one is their arguments.
In the NH case, $\tilde\mu_{NS_{2,3}}$ runs the whole ranges and there are no correlations between them.
In the IH case, $\tilde\mu_{NS_{2,3}}$ are respectively localized at nearby [130,225] and [170,300].

\section{Summary and Conclusions}
We have proposed an interesting assignment of positive modular weights for 4D fields that enables us to construct Inverse Seesaw and Linear Seesaw models without any additional symmetries.
These scenarios can be realized at a lower energy scale that would be attractive in view of verifiability via current experiments.
At first, we have discussed possibilities for the positive modular weights from a theoretical point of view.
Then, we have shown two examples; explicit constructions of IS and LS models,
and demonstrated some predictions via numerical analyses.
We find that the minimal realization of these models is achieved by positive modular weights for 4D superfields.

Before closing this paper, we would like to emphasize that
this possibility of positive modular assignment would possess more interesting aspects in model building and vast applications for flavor physics. 
The positive modular weights will give new insights not only into the flavor structure of quarks and leptons but also into higher-dimensional operators in the SM effective field theory~\cite{Kobayashi:2021pav,Kobayashi:2022jvy}.
 

\section*{Acknowledgments}
The work was supported by the Fundamental Research Funds for the Central Universities (T.~N.), JSPS KAKENHI Grant Numbers JP20K14477 (H. Otsuka), JP23H04512 (H. Otsuka) and JP23K03375 (T. K.).

\bibliography{MA4-QLDMmg2.bib}
\end{document}